%% file: paper.tex
\documentclass[a4paper,11pt]{article}
\pdfoutput=1
\pdfoutput=1 
\usepackage{amssymb}
\usepackage{graphicx}
\usepackage{jheppub}
\usepackage{lineno}
\usepackage{subfigure}
\usepackage{amsmath}
\usepackage{graphicx}
\usepackage{hyperref}
\hypersetup{colorlinks=true,urlcolor=blue}

\title{Importance and construction of features in identifying new physics signals with deep learning}

\author[a]{Chang-Wei Loh,}
\author[a]{Rui Zhang,}
\author[a]{Yong-Heng Xu,}
\author[a]{Zhi-Qiang Qian,}
\author[b]{Si-Cheng Chen,}
\author[a]{He-Yang Long,}
\author[a]{You-Hang Liu,}
\author[a]{De-Wen Cao,}
\author[a]{Wei Wang}
\author[a,1]{and Ming Qi}

\affiliation[a]{Nanjing University, 22 Hankou Road, Nanjing, Jiangsu, China}
\affiliation[b]{Nanjing University of Aeronautics and Astronautics, 29 Jiangjun Dadao, Nanjing, Jiangsu, China}

\emailAdd{qming@nju.edu.cn}

\abstract{Advances in machine learning have led to an emergence of new paradigms in the analysis of large data which could assist traditional approaches in the search for new physics amongst the immense Standard Model backgrounds at the Large Hadron Collider. Deep learning is one such paradigm. In this work, we first study feature importance ranking of signal-background classification features with deep learning for two Beyond Standard Model benchmark cases: a multi-Higgs and a supersymmetry scenario. We find that the discovery reach for the multi-Higgs scenario could still increase with additional features. In addition, we also present a deep learning-based approach to construct new features to separate signals from backgrounds using the ATLAS detector as a specific example. We show that the constructed feature is more effective in signal-background separation than commonly used features, and thus is better for physics searches in the detector. As a side application, the constructed feature may be used to identify any momentum bias in a detector. We also utilize a convolutional neural network as part of the momentum bias checking approach.}


\begin{document}
\maketitle
\flushbottom
\section{Introduction}

After the discovery of a light Higgs boson \cite{HiggsATLAS,HiggsCMS}, there is a general expectation for particles from physics beyond the Standard Model (BSM) to be observed in the ATLAS and CMS detectors at the Large Hadron Collider (LHC). However, none has been found so far. Often, to discriminate signals originating from new physics against the immense Standard Model (SM) backgrounds, various signal-background classification features, including raw kinematic features of final states physics objects as measured by the detector and derived features which can be computed from the combinations of these raw features, are used. Typical raw features include the transverse momentum $p_T$ and pseudorapidity $\eta$ of the final states physics objects and missing energy $MET$ of the physics events, while derived features are commonly physics inspired which includes the invariant masses of a set of physics objects. The choice of using any features lies in whether the classification performance and hence the discovery significance could be improved with the inclusion of a feature as part of the physics event selection cuts.

Employing machine learning in signal-background classification problems in particle physics is not new. In fact, machine learning approaches typically outperform the traditional linear cuts used for event selections, as the latter could not capture the non-linear correlations among the classification features. For instance, in the search for new physics in the $WWbb$ channel with the ATLAS detector, which assumed a multi-Higgs boson cascade decay \cite{multiHiggs} as a benchmark process, a machine learning tool, i.e. boosted decision tree (BDT) \cite{BDT} signal-background classifier has been used. In the continuous attempt to gain an edge in discovering new physics, a new machine learning paradigm, known as deep learning \cite{deeplearning} has recently caught the attention of the particle physics community.

Deep learning is a class of machine learning which has had ubiquitous success in a plethora of disparate fields, ranging from arts \cite{arts} and language \cite{language} to genetics \cite{genetics} and drug discovery \cite{biomedicine}. In high-energy physics, various studies have been done for its potential use in particle collisions \cite{CNNcollision}, exotics searches \cite{NatureBaldi}, jet classifications \cite{deepjet,deepjetqg,deepjetJHEP} and the monitoring of superconducting magnets at the LHC \cite{deepmonitoring}. Deep learning can be regarded as an extension of the artificial neural network, but with more hidden layers and more versatile in terms of the connection between neurons in the layers. The versatility of the connections enables the emergence of new network architectures such as the autoencoder \cite{autoencoder}, convolutional neural network \cite{CNN}, long short-term memory \cite{LSTM} and recursive neural network \cite{recursive}. Succinctly, deep neural networks (DNN) is a highly effective data-driven function approximator that seeks to model the quantity of interest $y$ using data from a vector of inputs $x$ with $DL(x,p) = y$, where $p$ are parameters of the DNN; their values are found during the training stage of the network. In the case of a signal-background classification problem, $x$ would be the classification features and $y$ would be the event class, i.e. a signal or a background. DNN is appealing as it bypasses the need for assumptions about the underlying mechanisms that produce the data, and instead, assist us in gaining insights as to the mechanisms that produce them.

In many areas, it is true that machines outperform humans on problems tailored to achieve a certain aim. Machines optimizes their inputs with one aim in mind, i.e. to achieve the best possible target pertaining to that aim. In particle physics, given kinematic inputs from the detectors, we construct detector-agnostic features with physics interpretation that may or may not work optimally for signal-background separations in an experimental setting. This contradicts the aim of optimizing a signal-background separation. To search for some new unknown physics, one desires to achieve the best possible discovery reach given the amount of data collected at the LHC. Could machine learning assist in constructing new features to achieve this aim optimally?

Common to all machine learning endeavors in physics event selection cuts, is the search for new classification features as inputs to the machine learning classifiers. For instance, in supersymmetry (SUSY) searches, much work has been done to craft new classification features \cite{stransverse,razor,razor2,superrazor} to assist in increasing the signal-background discrimination power. However, an increase in the number of features used in machine learning approaches would evoke a high computational cost in doing a single event class prediction and the curse of dimensionality; the latter refers to the decrease in the accuracy of a classifier prediction with increasing number of input features when the data is sparse in addition to the larger uncertainty in the prediction.

Therefore, in this work, we first compare deep learning with BDT, where the latter is a commonly used machine learning approach in particle physics, in their effectiveness on new physics searches. We also perform a feature importance ranking with respect to the machine learning methods to identify the subset of features that gives an optimal classification. We demonstrate this using two benchmark processes: multi-Higgs, which is also of relevance to models with an extended Higgs sector \cite{Higgsdoublet}, and SUSY. The datasets used here has been used previously in \cite{NatureBaldi}. We then present an approach based on deep learning to construct new features from some raw kinematic features obtained from a simulation of the ATLAS detector using the Higgs(Z boson) decaying to two tau leptons as the signal(background) class, where both decays have already been observed at the LHC \cite{HtautauATLAS,HtautauCMS}. Using such already known signals and backgrounds would provide a good testing ground at the LHC for the deep learning-based approach. We also discuss some implications of the constructed features, in particular as a momentum bias checking tool.

\section{Feature Importance in Beyond Standard Model Signals}
\subsection{Multi-Higgs Signal}

\begin{figure}[htbp]
	\centering
	\includegraphics[width=8.cm]{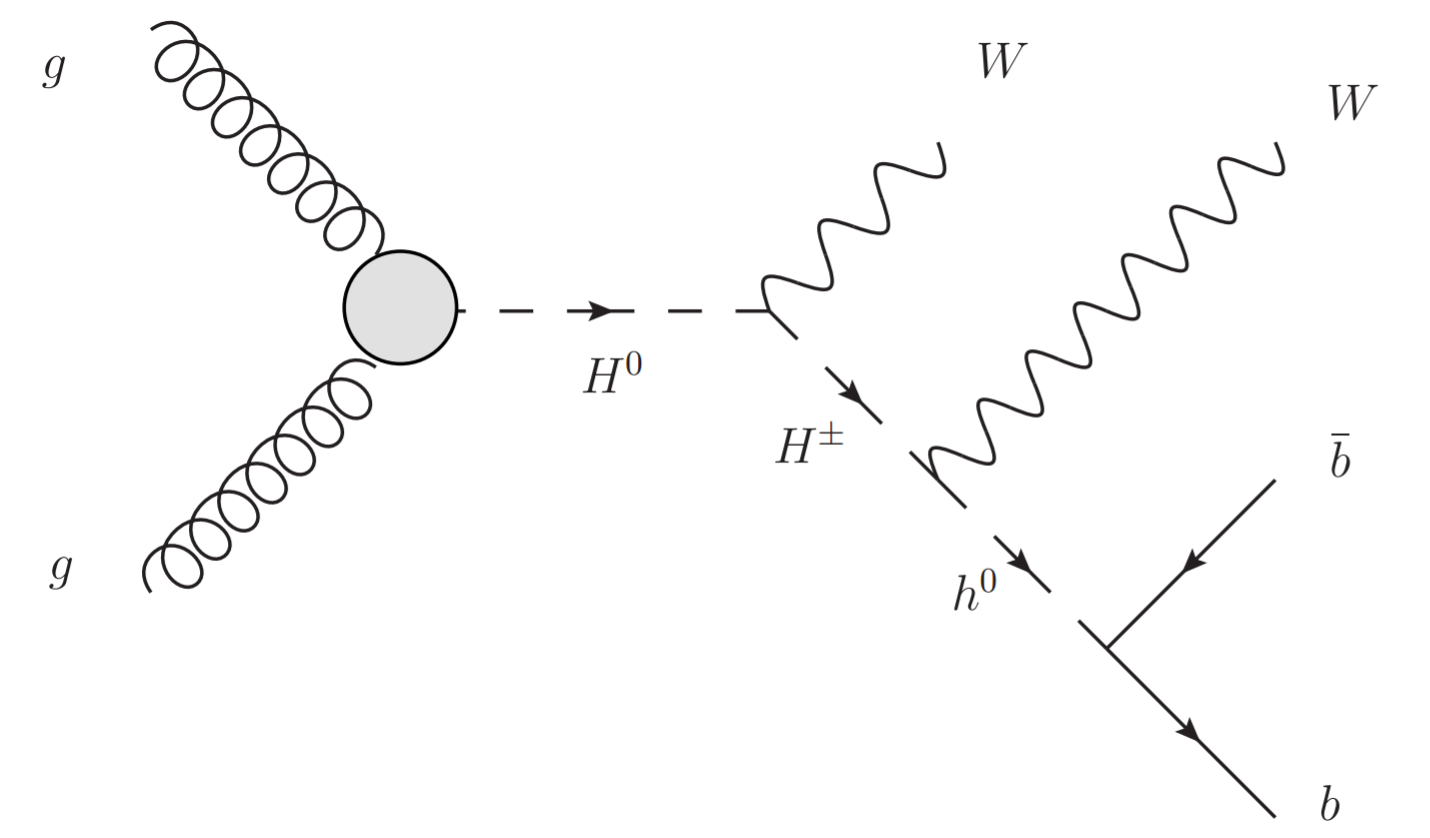}
	\caption{\label{Higgs_sig} Diagram showing the multi-Higgs cascade decay signal. }
\end{figure}

In the multiple Higgs bosons scenario, the model contains a heavy neutral Higgs $H$ (425 GeV), an intermediate charged Higgs $H^{\pm}$ (325 GeV) and a light Higgs boson $h$ (125 GeV). The signal is a gluon fusion process producing a cascade decay leading to $h$ (with diagram shown in Figure \ref{Higgs_sig}):
\begin{equation}
gg \rightarrow H^0 \rightarrow W^{\mp}H^{\pm} \rightarrow W^{\mp}W^{\pm}h \rightarrow jjl\nu b\bar{b},
\end{equation}
where the light Higgs decays into the $b\bar{b}$ channel.
The dominant background to this signal is a $t\bar{t}$ production. 
A total of 11 million Monte Carlo (MC) events are available for the signal and background. These MC events were produced with a Madgraph \cite{Madgraph} event generator for 8 TeV pp collisions at the LHC using Pythia \cite{Pythia} for parton showering and hadronization processes. The simulation of the ATLAS detector was performed with Delphes \cite{Delphes}.
A total of 26 features were considered for the machine learning approaches, wherein 19 of which were raw kinematic features from the final state physics objects, for instance $p_T$, $\eta$, $\phi$ of the lepton and jets, missing energy $MET$ and b-tag scores, and seven derived features based on the realization that the masses of the Higgs bosons, W and top quark can be computed from the final state physics objects. The following are the derived features:
\begin{itemize}
	\item $m_{WWbb}$: $WWb\bar{b}$ mass which is related to the heavy Higgs $H$;
	\item $m_{Wbb}$: $Wb\bar{b}$ mass which is related to the charged Higgs $H^{\pm}$;
	\item $m_{bb}$: $b\bar{b}$ mass which is related to the light Higgs $h$;
	\item $m_{j\ell\nu}$: mass of the lepton$+MET+$jet which is related to the top quark;
	\item $m_{jjb}$: tri-jet mass which is related to the top quark;
	\item $m_{jj}$: dijet mass which is related to the strongly decaying W;
	\item $m_{\ell\nu}$: mass of the lepton+$MET$ which is related to the weakly decaying W.
\end{itemize}

\begin{figure}
	\centering
	\includegraphics[width=12.cm]{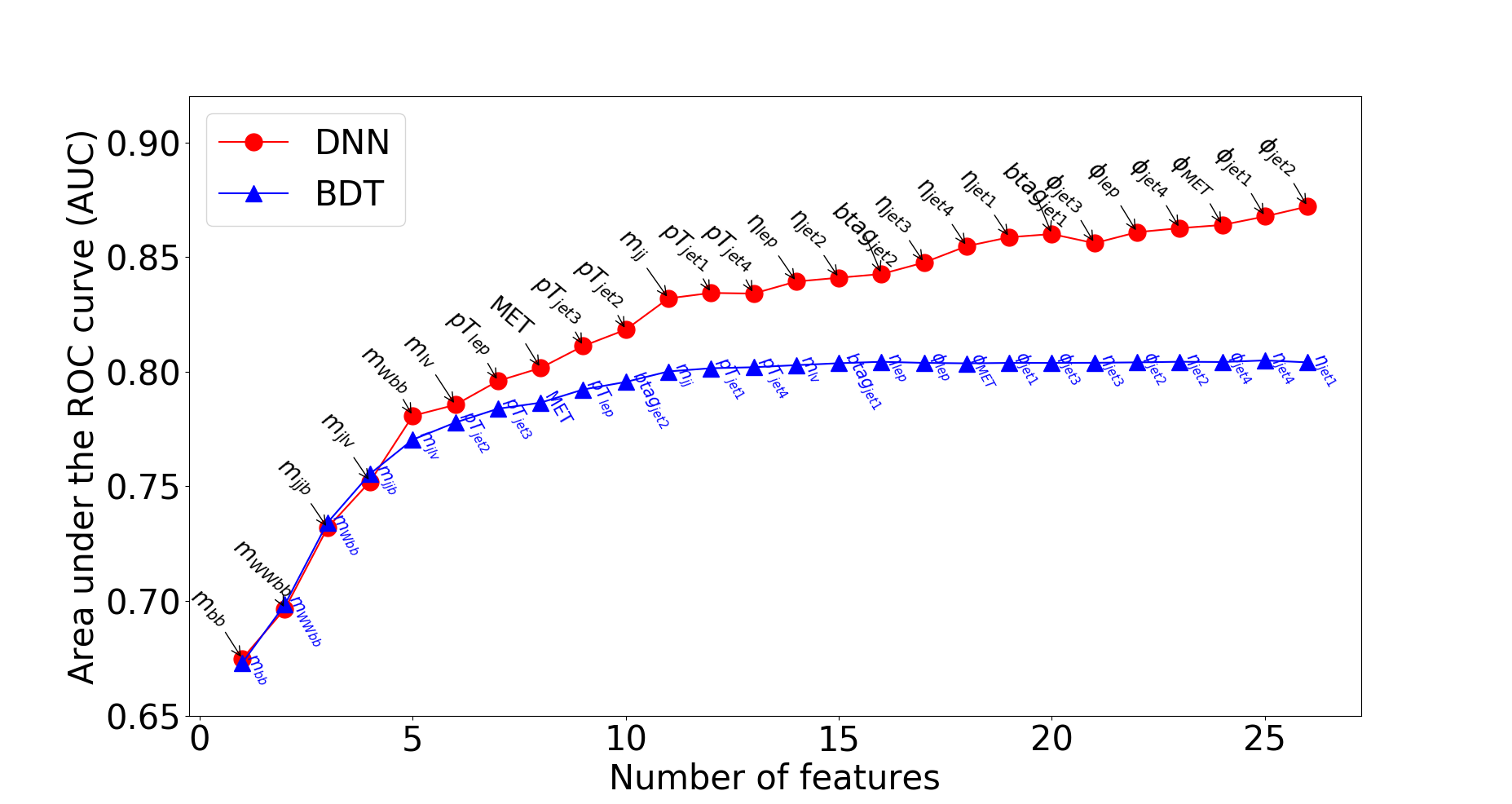}
	\caption{\label{Higgs_featureranking}   Feature importance ranking shown as the AUC vs. number of features $k$ using DNN and BDT. For each $k$-th data point, we also indicate the $k$-th important feature $F_k^*$ used to form the optimal feature set $\{F_k^*,...,F_2^*,F_1^*\}$ to obtain the corresponding AUC.}
\end{figure}

\begin{figure}
	\centering
	\includegraphics[width=10.cm]{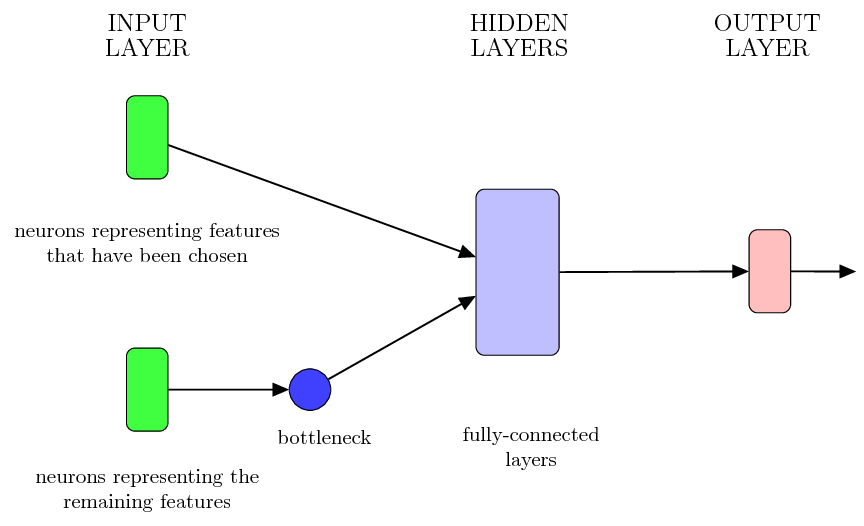}
	\caption{\label{architecture}   Architecture of the deep network with a bottleneck neuron used to obtain the feature importance ranking. }
\end{figure}

For ensuring the efficient training of the machine learning approaches, the MC data was preprocessed in order to standardize the features to some common range of values. This was done by scaling the features, i.e. dividing each feature by its maximum value within the dataset used. To measure the classification performance of machine learning approaches as well as simultaneously perform a feature importance ranking for the 26 features, we plotted the AUC, i.e. the area under the receiver operating characteristic curve (ROC) vs. the number of features $k$ used to obtain the AUC result (see Figure \ref{Higgs_featureranking}).
In brief, an ROC is a plot of the signal efficiency vs. background rejection. An AUC of 0.5 indicates random classification, while an AUC of 1 corresponds to a perfect classification.
In Figure \ref{Higgs_featureranking}, we show the classification performance and feature importance ranking for a deep learning and BDT classifier. For each $k$-th point in Figure \ref{Higgs_featureranking}, we indicate the $k$-th most important feature which was added to the subset of important features to obtain the AUC. For the BDT, the optimal set of $k$ was found through a permutation of the values of each feature \cite{permutation}. For deep learning, we designed a DNN architecture with a bottleneck neuron (see Figure \ref{architecture}) that was trained to minimize the cross-entropy loss function with a softmax activation function at the output layer.
To identify the first important feature out of the 26 features using the DNN with a bottleneck neuron, 26 input neurons representing the 26 features are forced to pass through the bottleneck neuron before connecting to the first fully-connected layer of the DNN during the training stage. Once the training ends, the weight of each input neuron at the bottleneck is extracted. The feature with the largest weight is regarded as the first important feature. Subsequently, in order to identify the second most important feature out of the remaining 25 features, a total of 25 input neurons representing the remaining 25 features are forced to pass through the bottleneck neuron before connecting to the first fully-connected layer, while the input neuron corresponding to the most important feature connects to the first fully-connected layer unhindered. In this manner, the DNN is encouraged to search for the best feature to be combined with the first important feature to minimize the cross-entropy of the DNN. Similar to the previous DNN training round, the weights of the 25 input neurons at the bottleneck are extracted when the training ends; the second most important feature is the one with the largest weight among the 25 features. Following this procedure, the $(k+1)$-th important feature $F_{k+1}^*$ will be selected from the remaining features which are not one of the $k$ important features $\{F_k^*,...,F_2^*,F_1^*\}$ found previously. To obtain the AUC for the set with $k$ features, the bottleneck is removed and the DNN is retrained with the $k$ features as input neurons.

\begin{figure}
	\centering
	\includegraphics[width=10.cm]{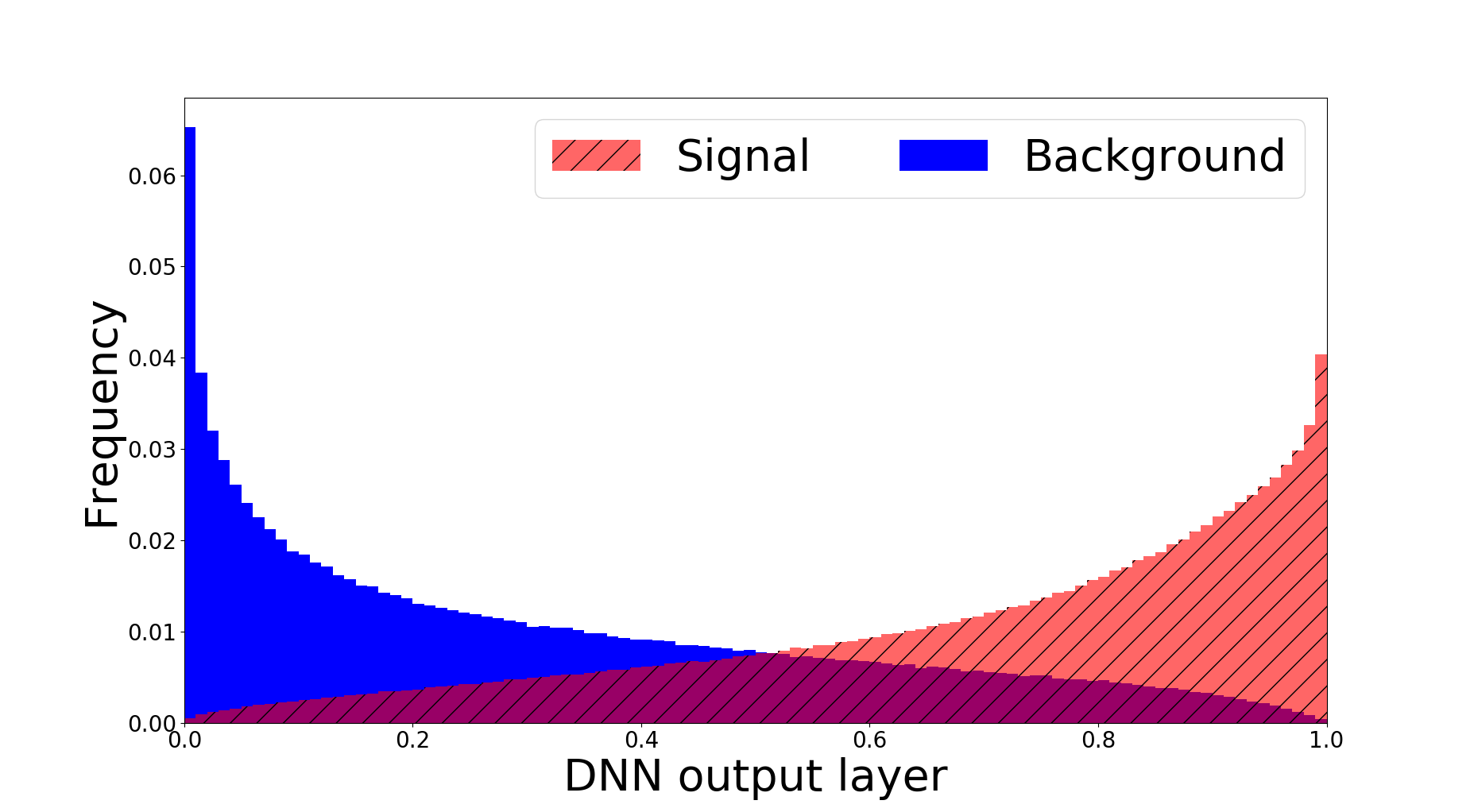}
	\caption{\label{Higgs_DNN_softmax}  Distribution of the signal and background events based on the softmax value of the output layer. }
\end{figure}
\begin{figure}[h]
	\centering
	\subfigure[$m_{bb}$] {
		\includegraphics[width=0.31\textwidth]{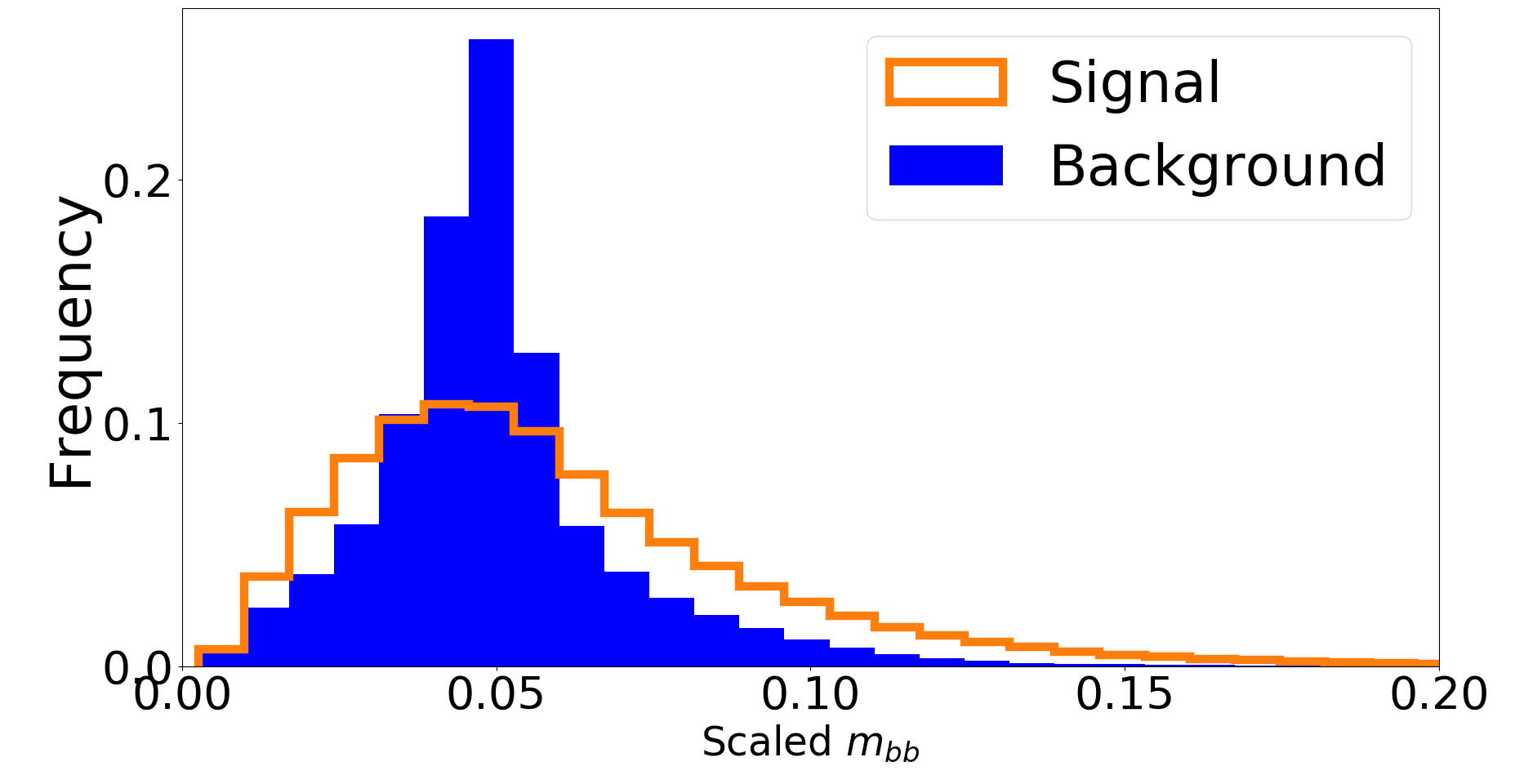}
		\label{F1}
	}
	\subfigure[$m_{WWbb}$] {
		\includegraphics[width=0.31\textwidth]{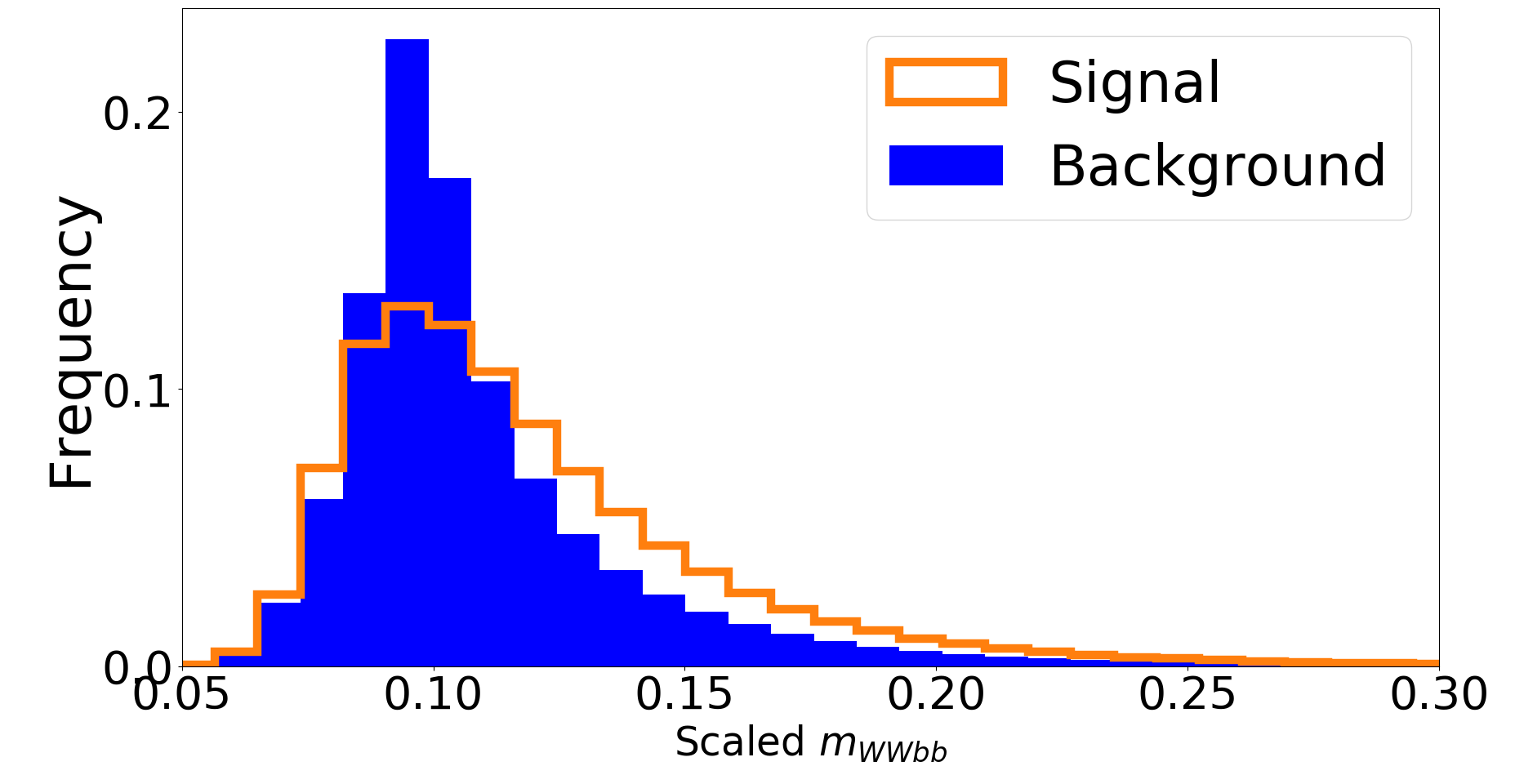}
		\label{F2}
	}
	\subfigure[$m_{jjb}$] {
		\includegraphics[width=0.31\textwidth]{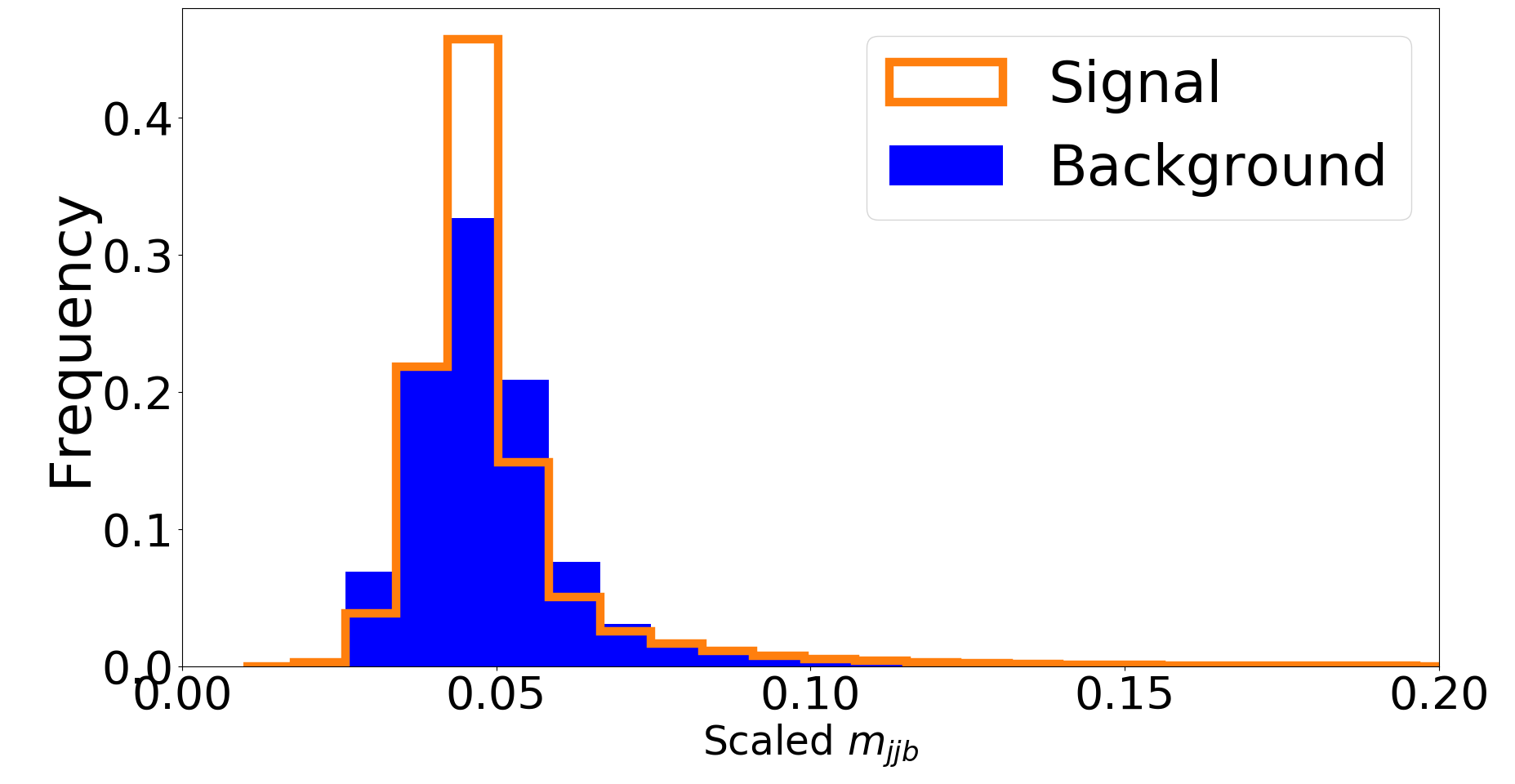}
		\label{F3}
	}
	\subfigure[$m_{j\ell\nu}$] {
		\includegraphics[width=0.31\textwidth]{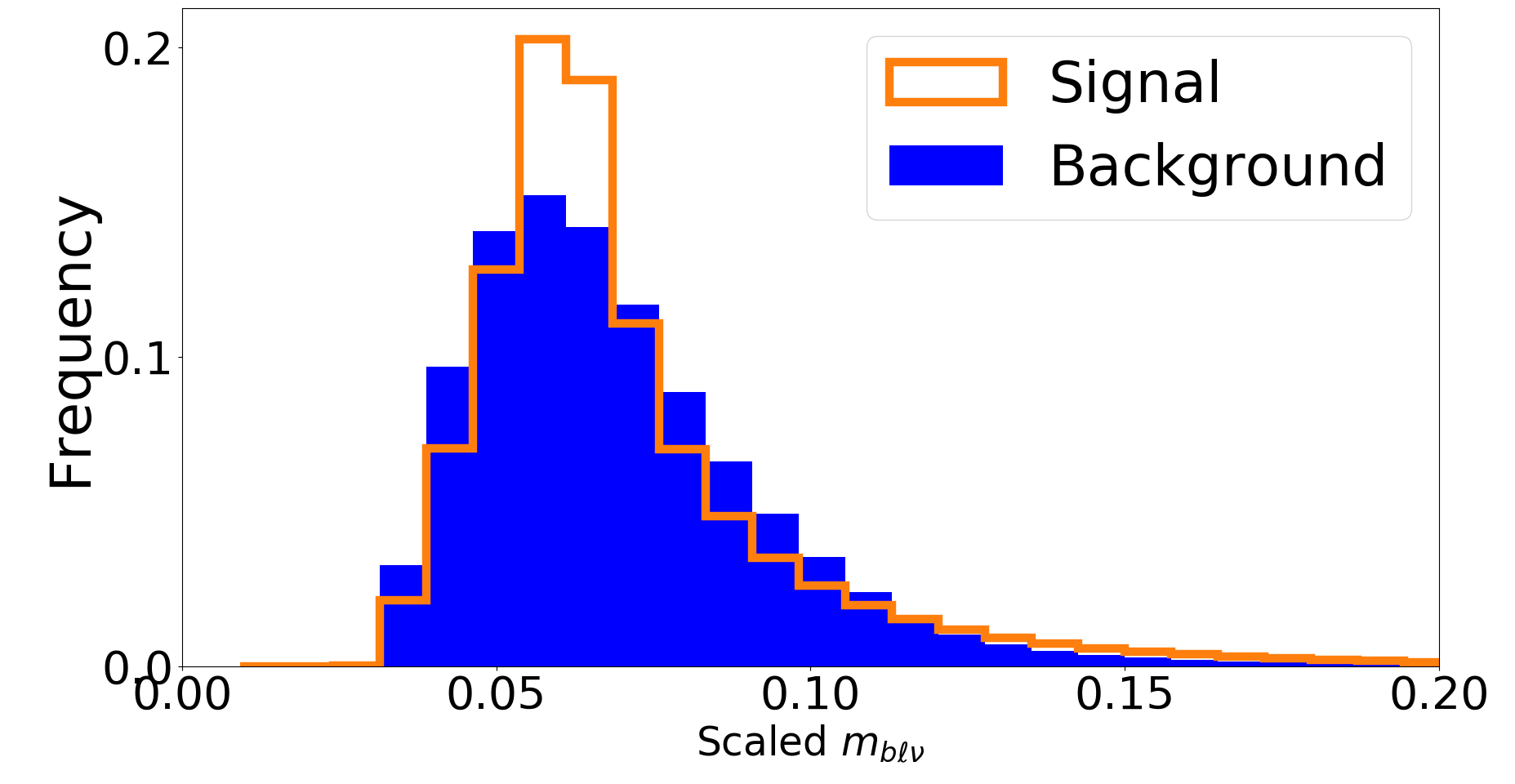}
		\label{F4}
	}
	\subfigure[$m_{Wbb}$] {
		\includegraphics[width=0.31\textwidth]{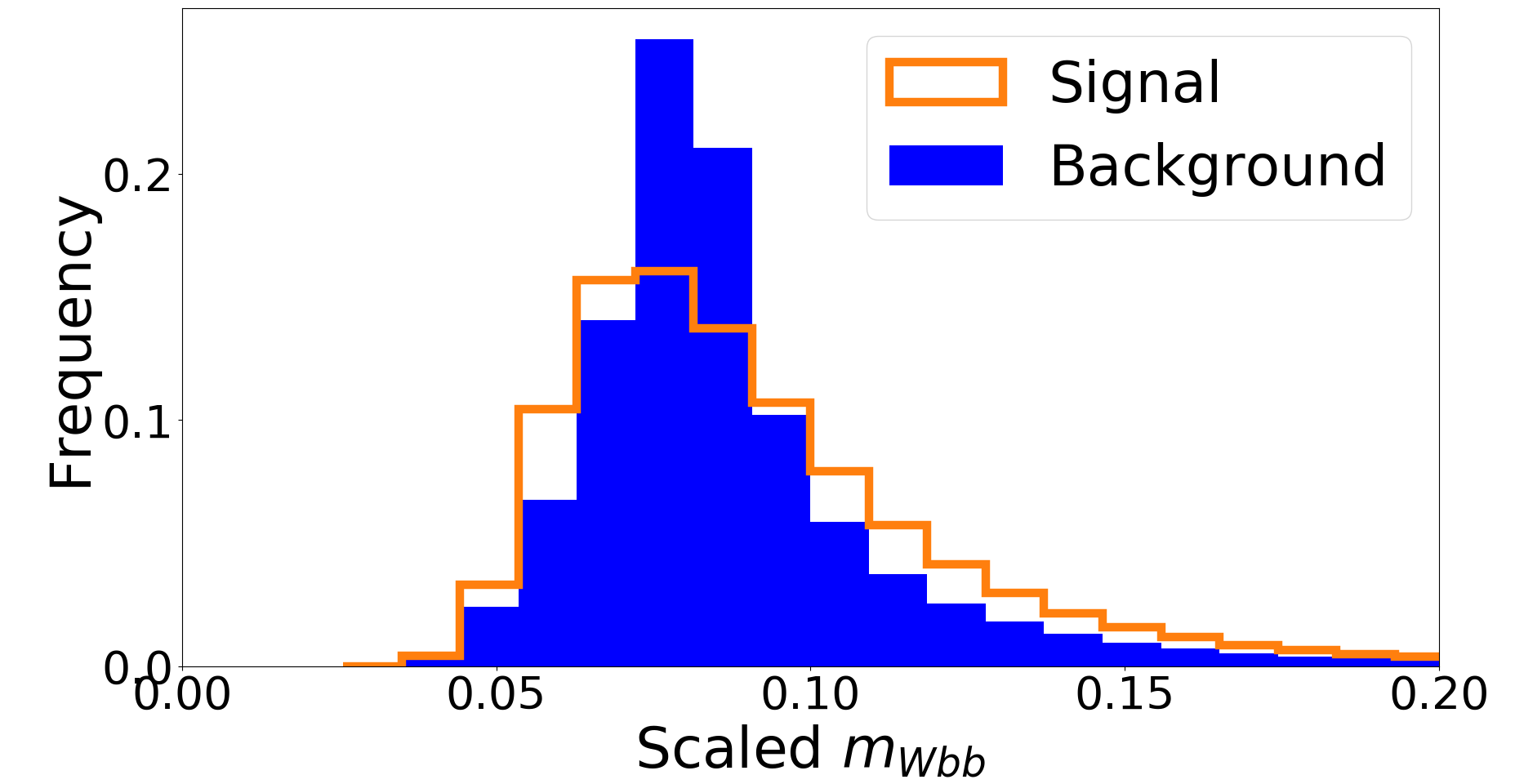}
		\label{F5}
	}
	\caption{Distributions of the 5 most important features with scaled ranges used by the DNN during training. }
	\label{Higgs_featuredist}
\end{figure}

It can be observed from Figure \ref{Higgs_featureranking} that when using the entire 26 features to classify the signal and background, DNN fares better than the BDT in line with the findings of \cite{NatureBaldi}. Figure \ref{Higgs_DNN_softmax} shows the distribution of the signal and background events based on the softmax value of the output layer of the DNN using all the 26 features as inputs, which could be interpreted as the probability of an event being a signal or otherwise when given the 26 features.
The capability of the DNN in exploiting the information contained within the features to increase the classification performance becomes visible compared to the BDT around when $k \geq 5$. The distributions of the set of 5 most important features with scaled ranges as used by the DNN during training are shown in Figure \ref{Higgs_featuredist}.
An observation one can make from Figure \ref{Higgs_featureranking} is that
the derived features did not use all the information contained within the raw features since adding new raw features after using all the derived features could still improve the classification performance. Also, the significance curve kept increasing with each additional feature, suggesting that the inclusion of further raw features could still help in improving the discovery reach.


\begin{figure}
	\centering
	\includegraphics[width=10.cm]{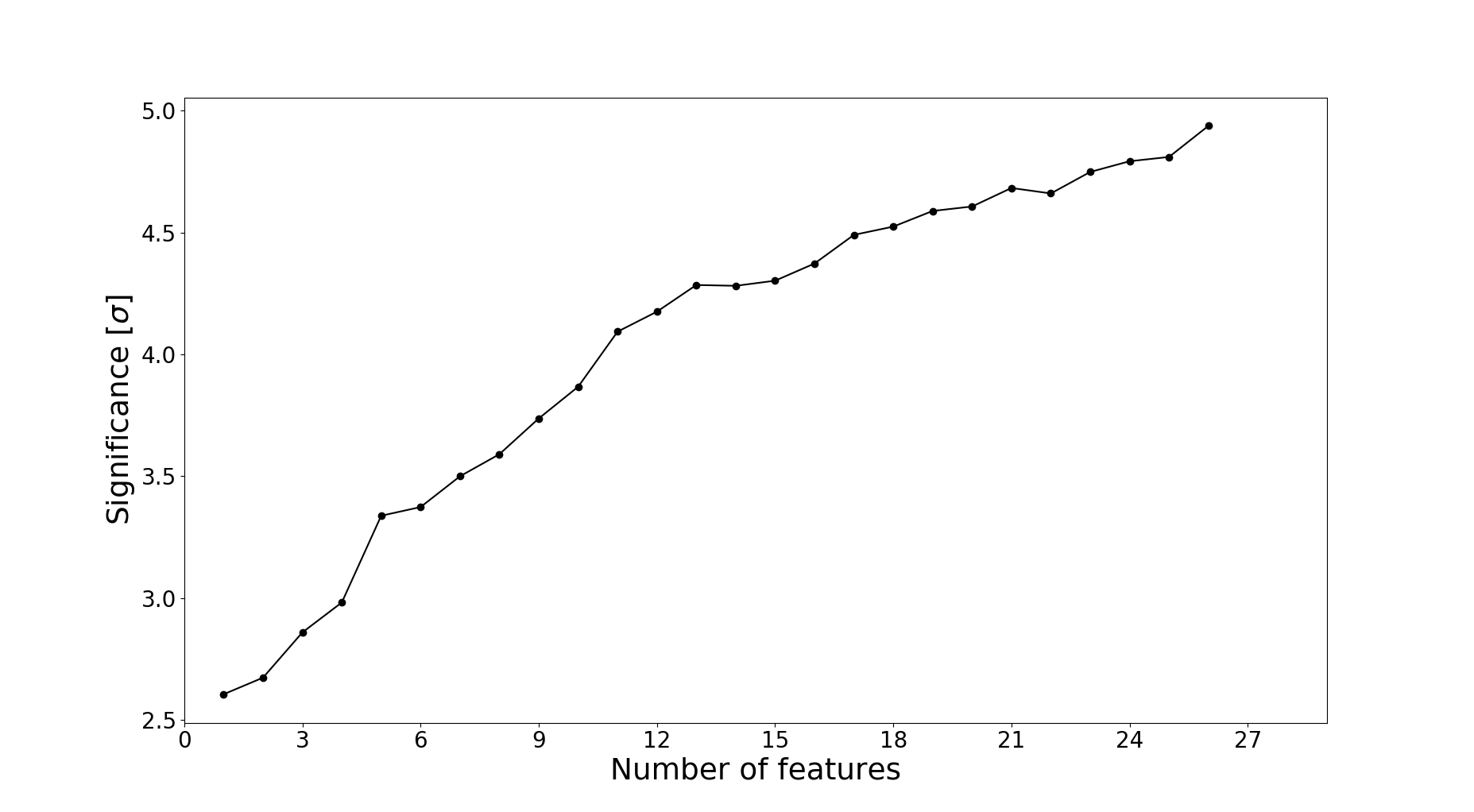}
	\caption{\label{Higgs_significance}  Significance (in units of number of standard deviations $\sigma$) vs. number of features for the multi-Higgs signal using DNN.}
\end{figure}

For the full set of 26 features, a 4.9 $\sigma$ significance is reachable assuming 80 signal and 1000 background events. Figure \ref{Higgs_significance} shows a plot of the significance $Z$ in units of $\sigma$ for a discovery reach vs. the number of features assuming the afore-mentioned number of signals and backgrounds under the background-only hypothesis, where $Z$ is as defined in \cite{Cowan}, i.e.
\begin{equation}
Z = \sqrt{2\Big( (S+B) \ln \Big( \frac{S+B}{B} \Big) - S\Big)},
\end{equation}	
which approaches $S/\sqrt{B}$ when $B \gg S$, where $S$ and $B$ are the number of signal and background events. From Figure \ref{Higgs_significance}, one can see that with additional features, there could still be room for improvement in the significance.

Naively, one would think that to optimally classify the events, one could resort to measuring the information "overlap" between the features, and select those which have the least overlaps in order to maximize the information coverage. However, this is not the case since we are interested in the gain in our knowledge about the event class $C$ (signal or background) given the relevant information contained in a feature or a subset of features, and hence it is only the information between the features which are relevant to $C$ that should be considered. More formally, by using the notion of the information entropy $H$, the reduction in the information entropy of $C$, denoted by $\Delta H$, when using information from an additional feature $F_{k+1}$ apart from already having information from a list of $k$ features, i.e. $F_k,...,F_2,F_1$ can be written as:
\begin{align}
\Delta H &= H(C | F_{k+1},F_k,...,F_2,F_1) - H(C | F_k,...,F_2,F_1) \label{eqn2} \\
&= H(F_{k+1} | C, F_k,...,F_2,F_1) - H(F_{k+1} | F_k,...,F_2,F_1), \label{eqn3} %
\end{align}
where Equation \ref{eqn3} can be obtained through algebraic manipulations of Equation \ref{eqn2} using the chain rule of the conditional information entropy.
Although the second term in Equation \ref{eqn3} is related to the correlation between the classification features, the first term is not as it depends also on the event class $C$. This clearly implies that an optimal classification of events, from an information-theoretic viewpoint, is only partially dependent on the correlations between classification features. In fact, $\Delta H$ can be interpreted as equivalent to the net information about $F_{k+1}$ that can only be gained through knowing $C$ and not by any of the other features $F_k,...,F_2,F_1$.
\begin{figure}
	\centering
	\includegraphics[width=13.cm]{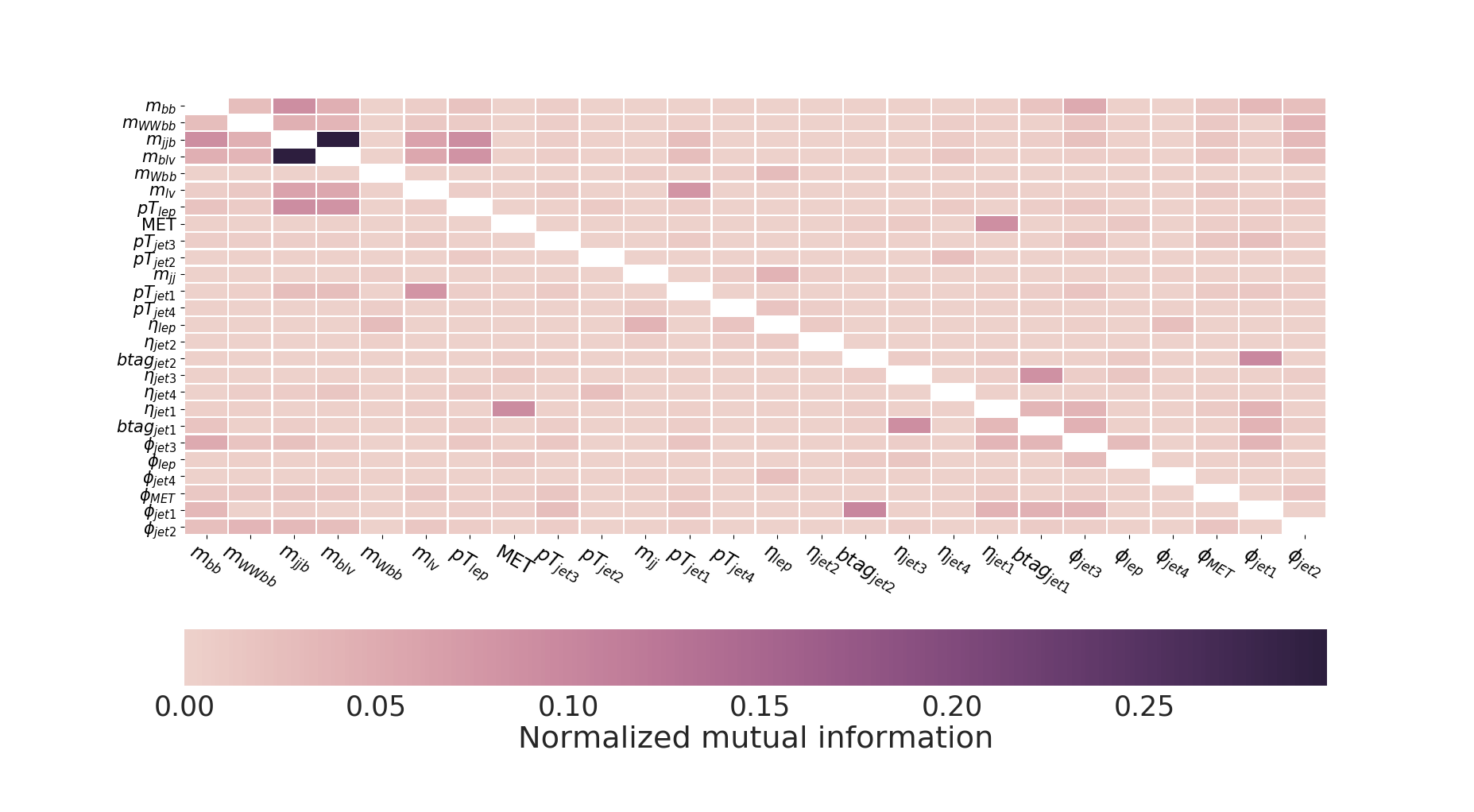}
	\caption{\label{mutual_info} Normalized mutual information $NMI$ between the features. The color of the heatmap shows the strength of the $NMI$. The normalized $NMI(F_i,F_i)$ are "whitened" for clearer visibility of the other $NMI$ values. The features have been arranged in accordance with the feature importance ranking from the DNN in Figure \ref{Higgs_featureranking}.}
\end{figure}
Figure \ref{mutual_info} shows the correlation between features using a normalized version of the mutual information $MI$, i.e. $MI(F_i,F_j)/\sqrt{MI(F_i,F_i) MI(F_j,F_j)}$. The features on both the x and y-axes are arranged according to the feature importance ranking as found by the DNN (see Figure \ref{Higgs_featureranking}). One can observe that there is no clear trend among the features in terms of their mutual information.


\subsection{Supersymmetry Signal}

\begin{figure}%
	\centering
	\includegraphics[width=.5\textwidth]{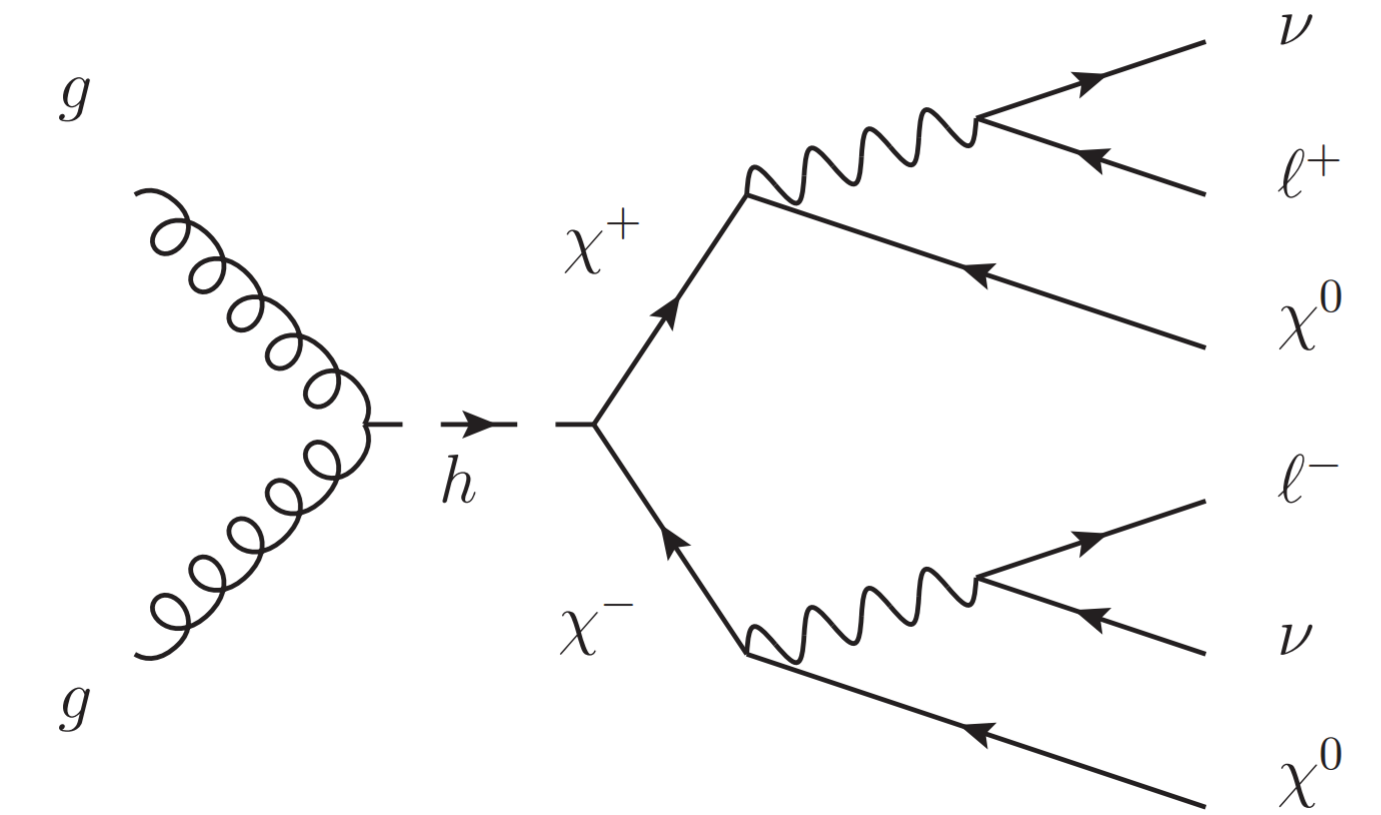}
	\caption{\label{SUSY_sig} Diagram showing the SUSY decay signal. }
\end{figure}

Here, we repeat the same procedure as in the multi-Higgs scenario, but this time with a SUSY benchmark process. The SUSY model is a gluon fusion process producing a pair of charged neutralinos $\chi^{\pm}$ (200 GeV); each then decays to a W boson and a neutral neutralino $\chi^0$ (100 GeV) acting as the lightest SUSY particle which travels through the detector as missing energy. This process can occur in various models including the Minimal Supersymmetric Standard Model (MSSM). Our signal is the channel with the final decay products: $l\nu \chi^0 l\nu \chi^0$ (see Figure \ref{SUSY_sig}).

The main background is a Standard Model $WW$ production with missing energy due to the two neutrinos from the decays of the $W$ bosons. The generation of the signal and background events are similar to that in the multi-Higgs case. A total of 18 features are considered here; 8 raw features and 10 derived features. The derived features are the stransverse mass \cite{stransverse}, razor \cite{razor,razor2} and super-razor \cite{superrazor} quantities. More details on the derived features can be found in \cite{NatureBaldi}.

\begin{figure}
	\centering
	\includegraphics[width=12.cm]{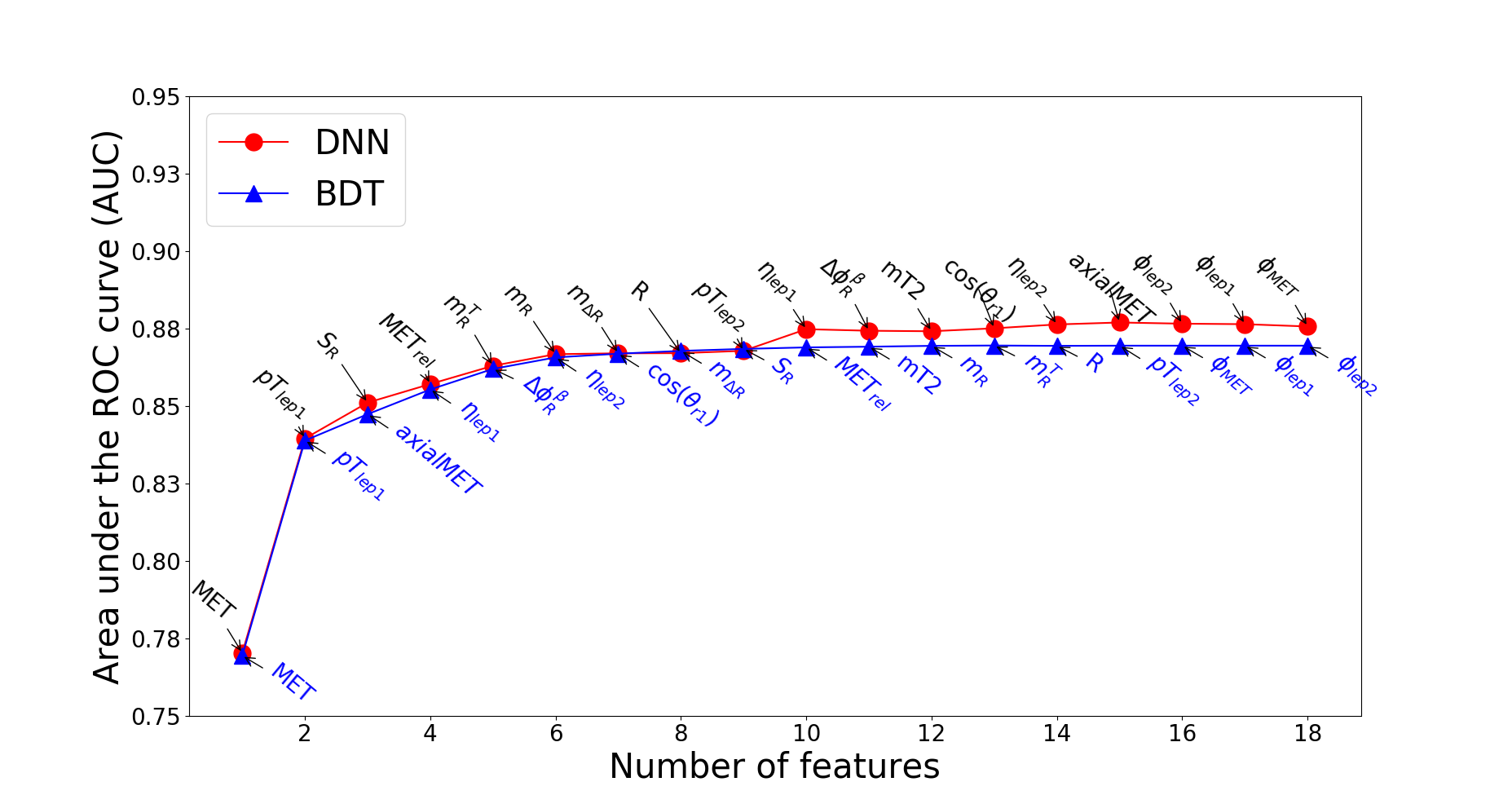}
	\caption{\label{SUSY_featureranking}   Feature importance ranking in terms of the obtained optimal AUC vs. number of features using DNN and BDT for the SUSY scenario.}
\end{figure}

Figure \ref{SUSY_featureranking} shows the feature importance ranking using DNN and BDT.
Similar to the multi-Higgs scenario, the DNN outperforms the BDT. However, in contrast with the multi-Higgs scenario, the classification performance with about ten optimal features for the SUSY case has already reach a plateau, i.e. adding additional features does not help the signal-background classifiers any further. As such, one needs to rely more on increasing the integrated luminosity of the $pp$ collisions to improve the discovery reach for this SUSY scenario. From Figure \ref{SUSY_featureranking}, one can see that the DNN classifier picked $MET$ as the most important feature, which should be expected since there is a large amount of $MET$ originating from the neutralinos $\chi^0$ in comparison with the lesser amount of $MET$ from the massless neutrinos in the SM background process. Perhaps what is surprising is that the second most important feature to be combined with $MET$ to form a two-feature set is a raw feature, i.e. the $p_T$ of a lepton. This suggests that the derived features did not entirely capture the information contained within the raw feature, and that the $p_T$ of the lepton offers valuable information not contained within the $MET$. It would be of interest to understand as to why this is the case, and perhaps new insights can emerge from the results here which can be used to derive better classification features.

\begin{figure}
	\centering
	\includegraphics[width=10cm]{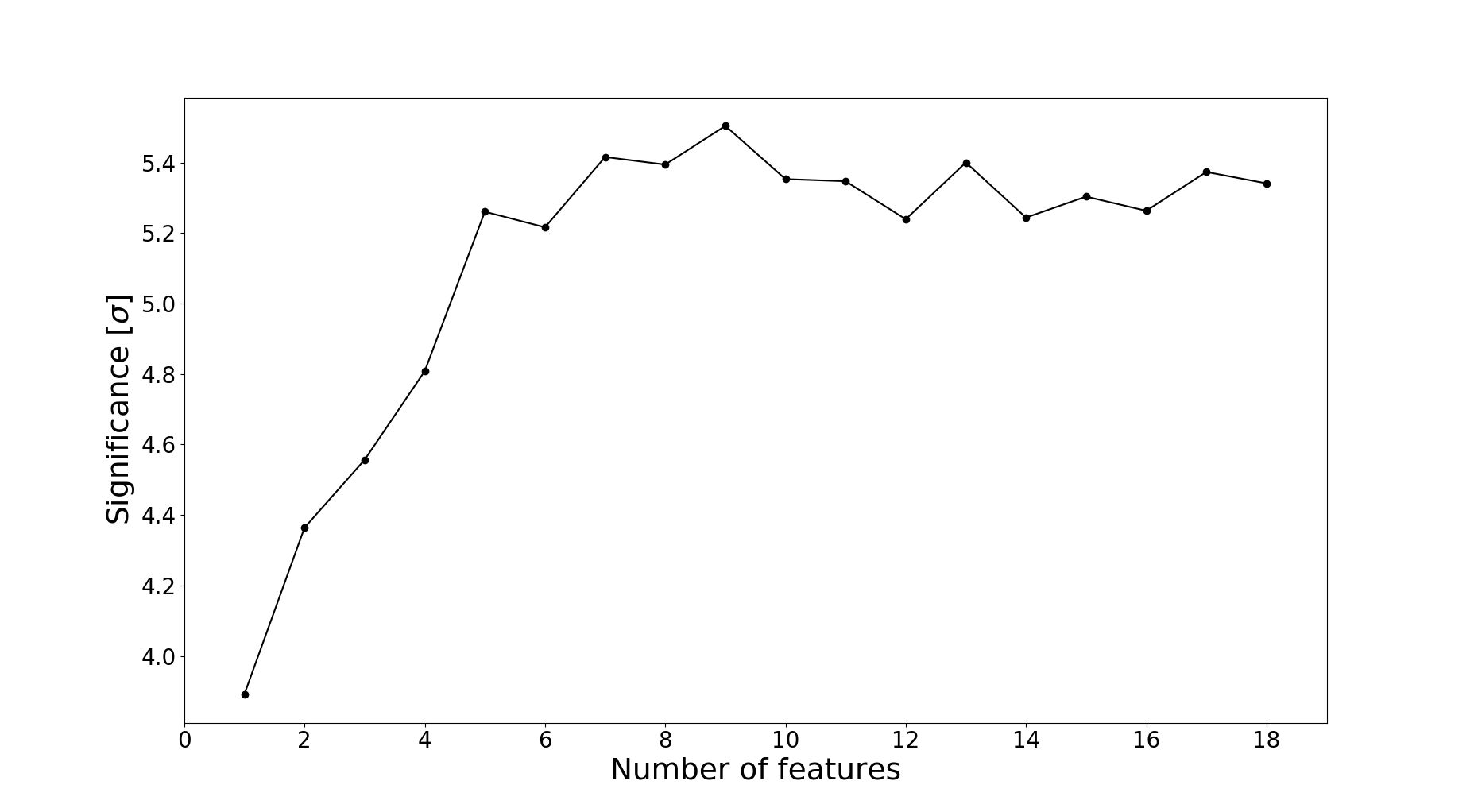}
	\caption{\label{SUSY_significance}  Significance vs. number of features for the SUSY signal obtained using DNN.}
\end{figure}

\begin{figure}
	\centering
	\includegraphics[width=10cm]{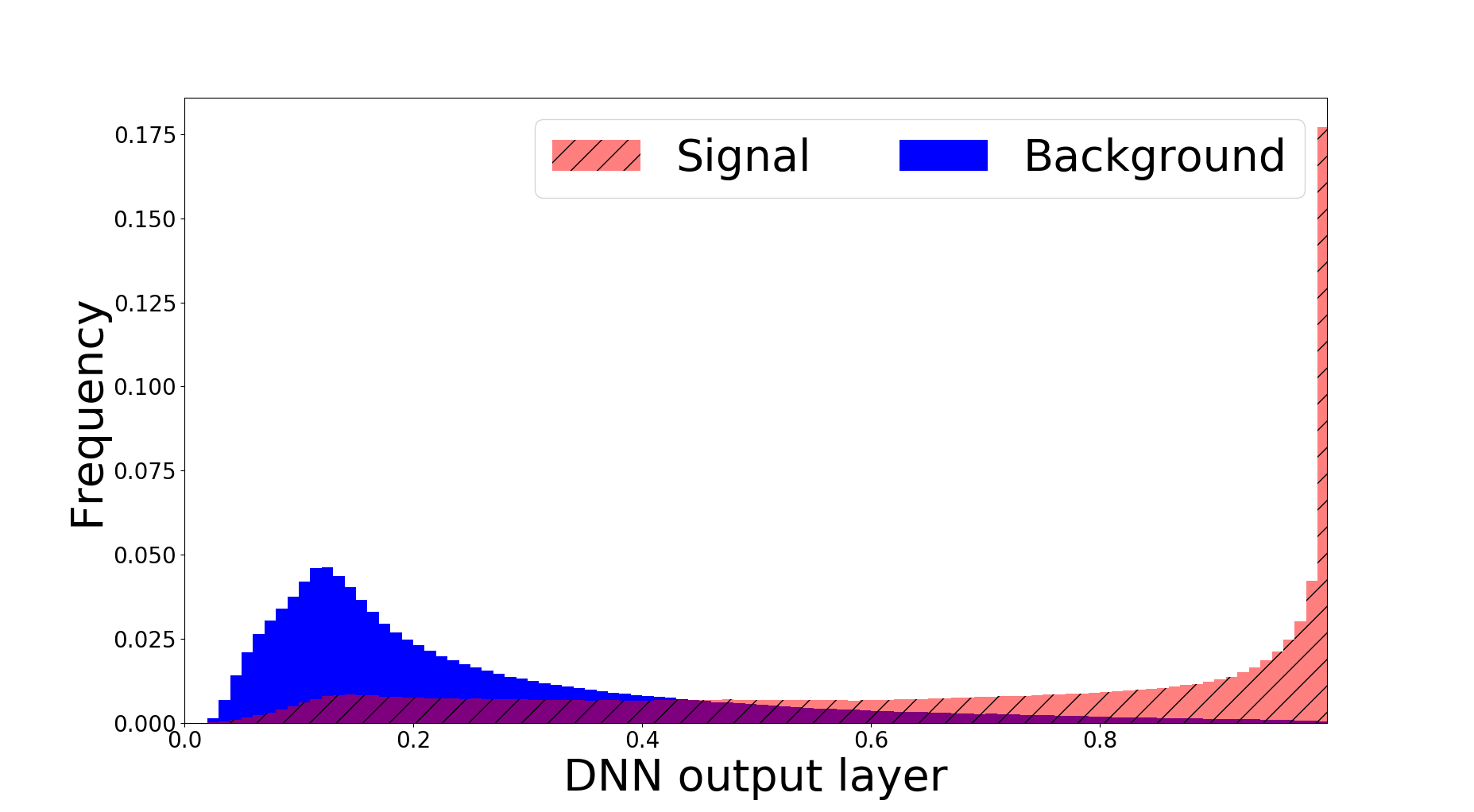}
	\caption{\label{SUSY_DNN_softmax}  Distribution of the SUSY signal and background events based on the softmax value of the output layer. }
\end{figure}

Figure \ref{SUSY_significance} shows a plot of the significance $Z$ in units of $\sigma$ for a discovery reach, assuming 40 signal and 1000 background events under the background-only hypothesis with a DNN.
With the set of ten most important features identified from the AUC curve, more than $5 \sigma$ significance could be reached. In line with the AUC vs. number of features curve, the significance curve has also reached a plateau. Performance curves like these are important to assist physicists in identifying the possible means of increasing the discovery potential of new physics. The distribution of the softmax value of the DNN output layer for the ten most important features is shown in Figure \ref{SUSY_DNN_softmax}.

\section{Constructing Features with Deep Learning}

The results in the previous Section show that the derived features in both multi-Higgs and SUSY scenario did not entirely capture the information content in the raw features. This motivates us to derive new classification features which could encapsulate more information content within the raw features to increase the physics discovery reach, and reduce the necessity of including a large number of features to obtain the same level of significance. Without including a large list of features into a signal-background classifier, one will also evade the curse of dimensionality. To this end, we explore the usage of deep neural networks to construct new features from the perspective of optimizing the signal-background discrimination strength of a classifier. The 4-momenta will be used as a demonstration hereafter.

One might surmise \textit{a priori} that the invariant mass of some final state physics objects would be a good classification feature. However, within an experimental setting where detector effects need to be taken into account, a modification to such a feature should be considered instead. To put it simply, if the kinematics data gathered from a detector is detector-based, then a classification feature should not be detector-agnostic in order to achieve an optimal signal-background separation.

\begin{figure}
	\centering
	\includegraphics[width=15cm]{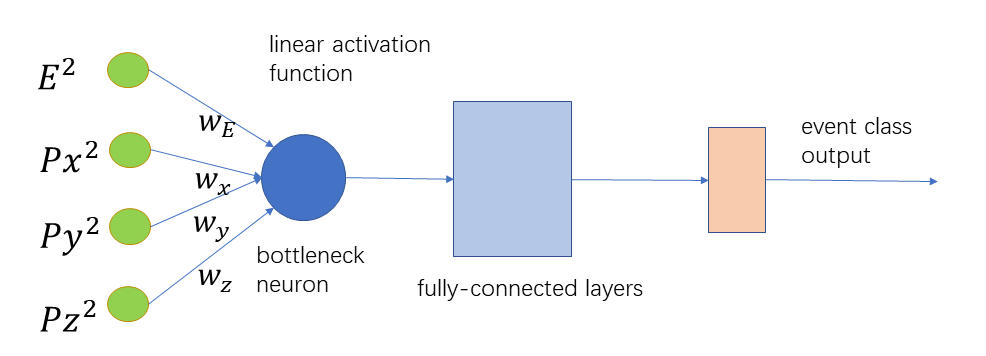}
	\caption{\label{architecture_featurecrafting}   Architecture of the deep network with a bottleneck neuron used for constructing new classification features. }
\end{figure}

Inspired by the invariant mass, let the following be a generic classification feature $F$ that we wish to construct:
\begin{equation}
F := w_E E^2 + w_x p_x^2 + w_y p_y^2 + w_z p_z^2,
\label{coeff_mass}
\end{equation}
where each $w_i$ is a detector-based coefficient. By interpreting the $w_i$ as the weights of the $(E^2,p_x^2,p_y^2,p_z^2)$ input neurons and using the optimization of the event classification as an aim for a purposefully designed DNN as shown in Figure \ref{architecture_featurecrafting}, these detector-based $w_i$ can be found.
The DNN architecture is similar to that of the one in Figure \ref{architecture}. However, the input neurons are $(E^2,p_x^2,p_y^2,p_z^2)$ which will pass through a bottleneck neuron such that the DNN would construct a classification feature in the form as defined in \ref{coeff_mass} that would optimize the separation of the signal and background.

For concreteness, we consider the classification problem of the signal: $h \to \tau\tau \to \ell\nu\ell\nu$ and background: $Z \to \tau\tau \rightarrow \ell\nu\ell\nu$. Such a classification problem can be easily implemented at the LHC since both signal and background events have been observed. Hence, this allows one to experimentally test this deep learning-based feature construction approach. Using the two leptons from the tau decays, we can construct an invariant mass classification feature $m_{\ell\ell}$:
\begin{equation}
m_{\ell\ell}^2 := (E_1 + E_2)^2 - (p_{x_1} + p_{x_2})^2 - (p_{y_1} + p_{y_2})^2 - (p_{z_1} + p_{z_2})^2,
\label{theory_mass}
\end{equation}
where the indices 1 and 2 corresponds to the first and second lepton in an event. Comparing Equation \ref{coeff_mass} and \ref{theory_mass}, the $w_i$ coefficients would be (1,-1,-1,-1) for $m_{\ell\ell}^2$.
Using simulated kinematics data for the ATLAS detector \cite{Htautau} to train the DNN in Figure \ref{architecture_featurecrafting} to minimize the cross-entropy loss function, we obtained the following classification feature $F_{\ell\ell}$:
\begin{equation}
F_{\ell\ell} := (E_1 + E_2)^2 + 5(p_{x_1} + p_{x_2})^2 + 5(p_{y_1} + p_{y_2})^2 - (p_{z_1} + p_{z_2})^2.
\label{F_sim}
\end{equation}
Note that, after the training of the DNN, the numerical values of the $w_i$ have been rounded to one significant figure to obtain $F_{\ell\ell}$.

\begin{figure}[h]
	\centering
	\subfigure[DNN] {
		\includegraphics[width=0.46\textwidth]{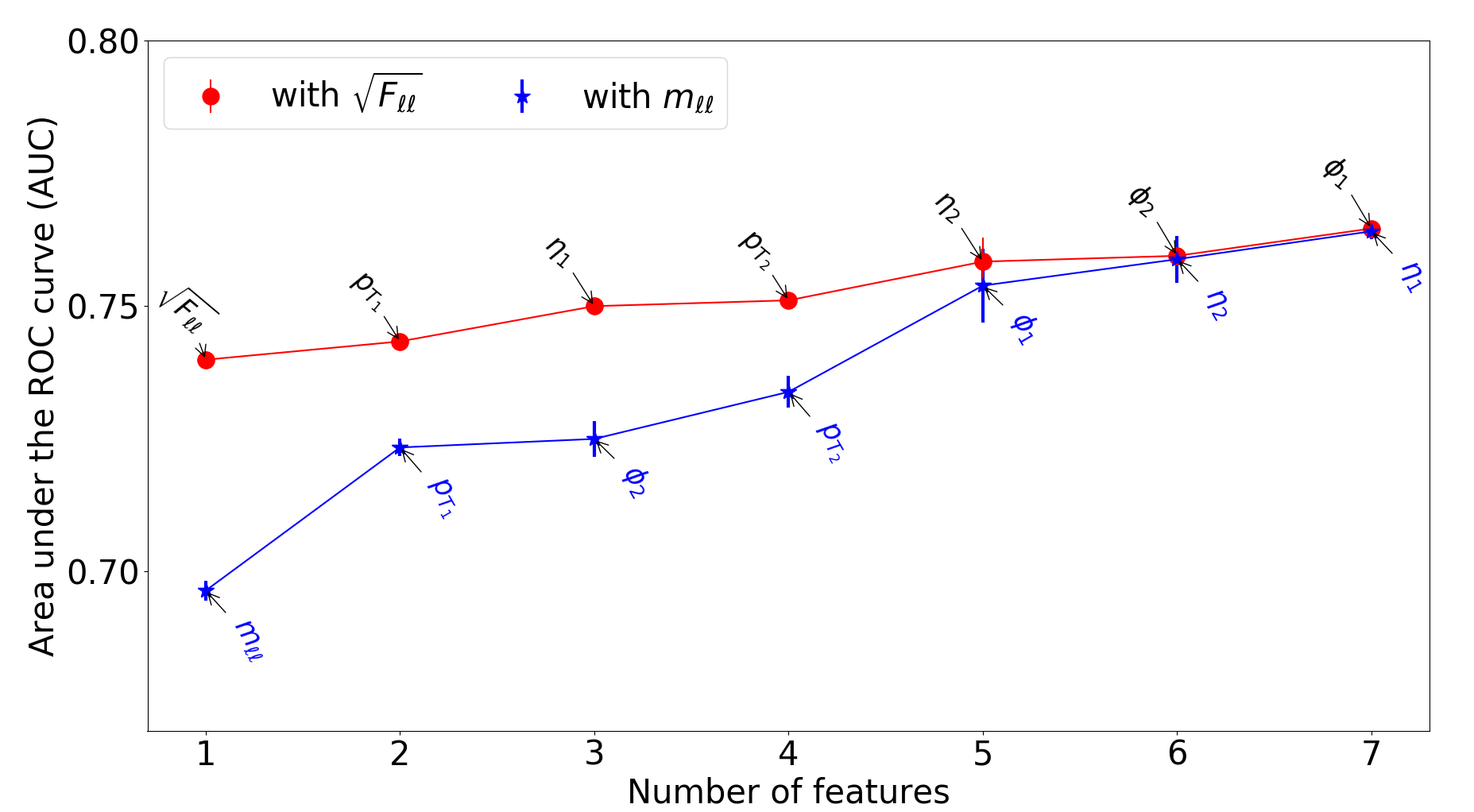}
		\label{mll}
	}
	\subfigure[BDT] {
		\includegraphics[width=0.46\textwidth]{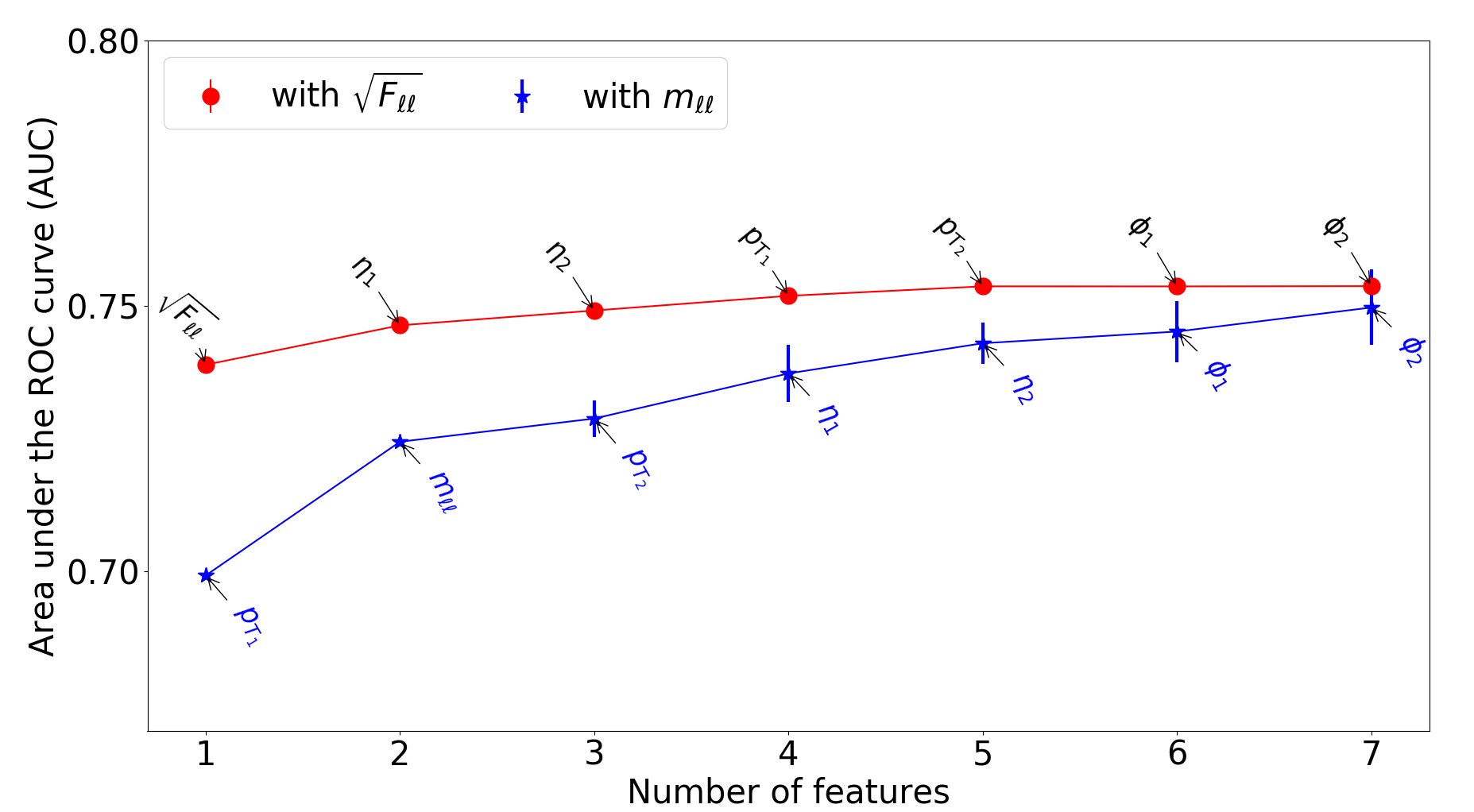}
		\label{Fsim}
	}
	\caption{Feature importance ranking and the classification performance of the feature set containing $\sqrt{F_{\ell\ell}}$ and the six raw features $\{p_{T_1},\eta_1,\phi_1,p_{T_2},\eta_2,\phi_2\}$, and another feature set containing $m_{\ell\ell}$ and the same six raw features when using (a) DNN and (b) BDT.}
	\label{AUC_with_constructedmass}
\end{figure}

\begin{figure}[h]
	\centering
	\subfigure[$m_{\ell\ell}$] {
		\includegraphics[width=0.46\textwidth]{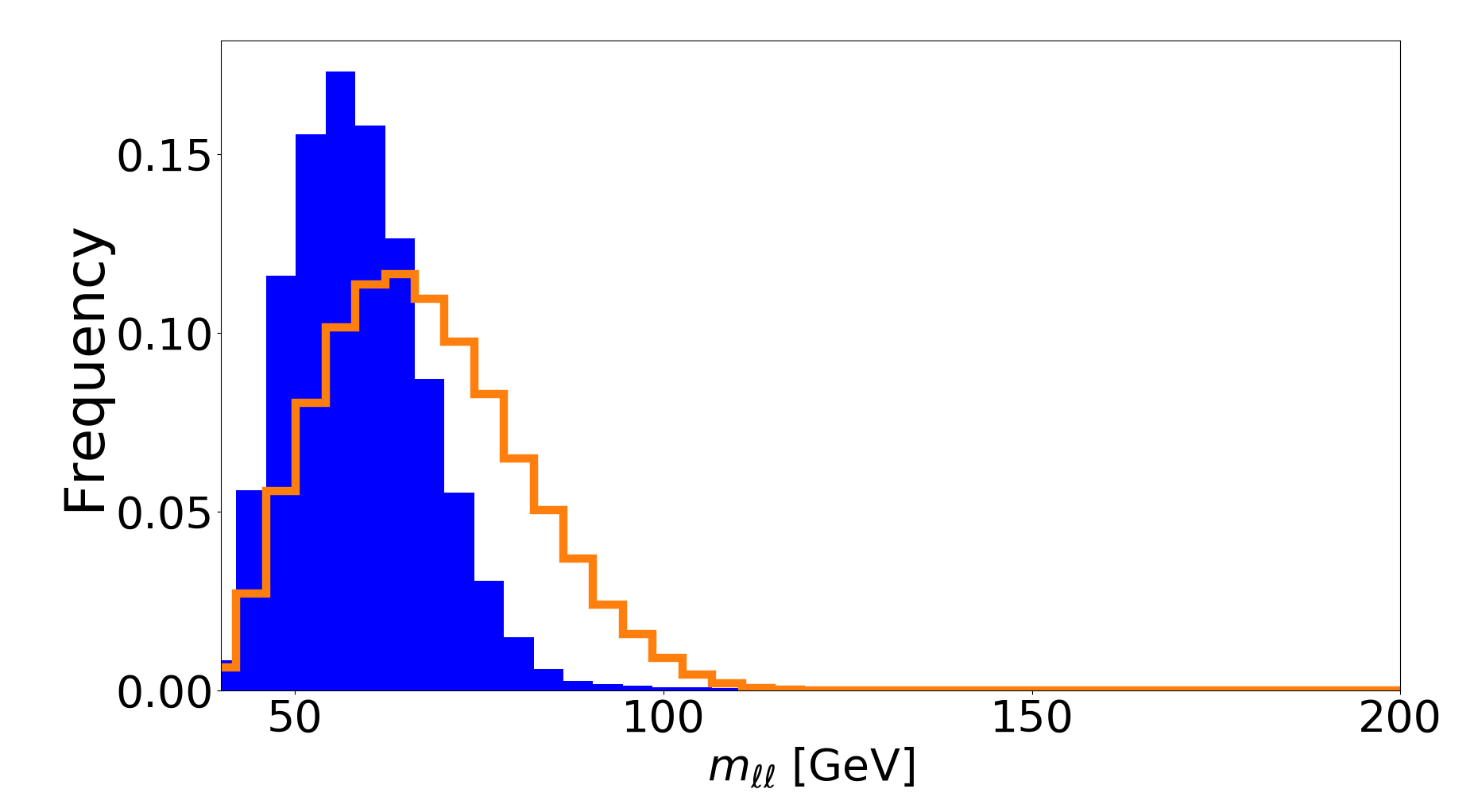}
		\label{mll}
	}
	\subfigure[$\sqrt{F_{\ell\ell}}$] {
		\includegraphics[width=0.46\textwidth]{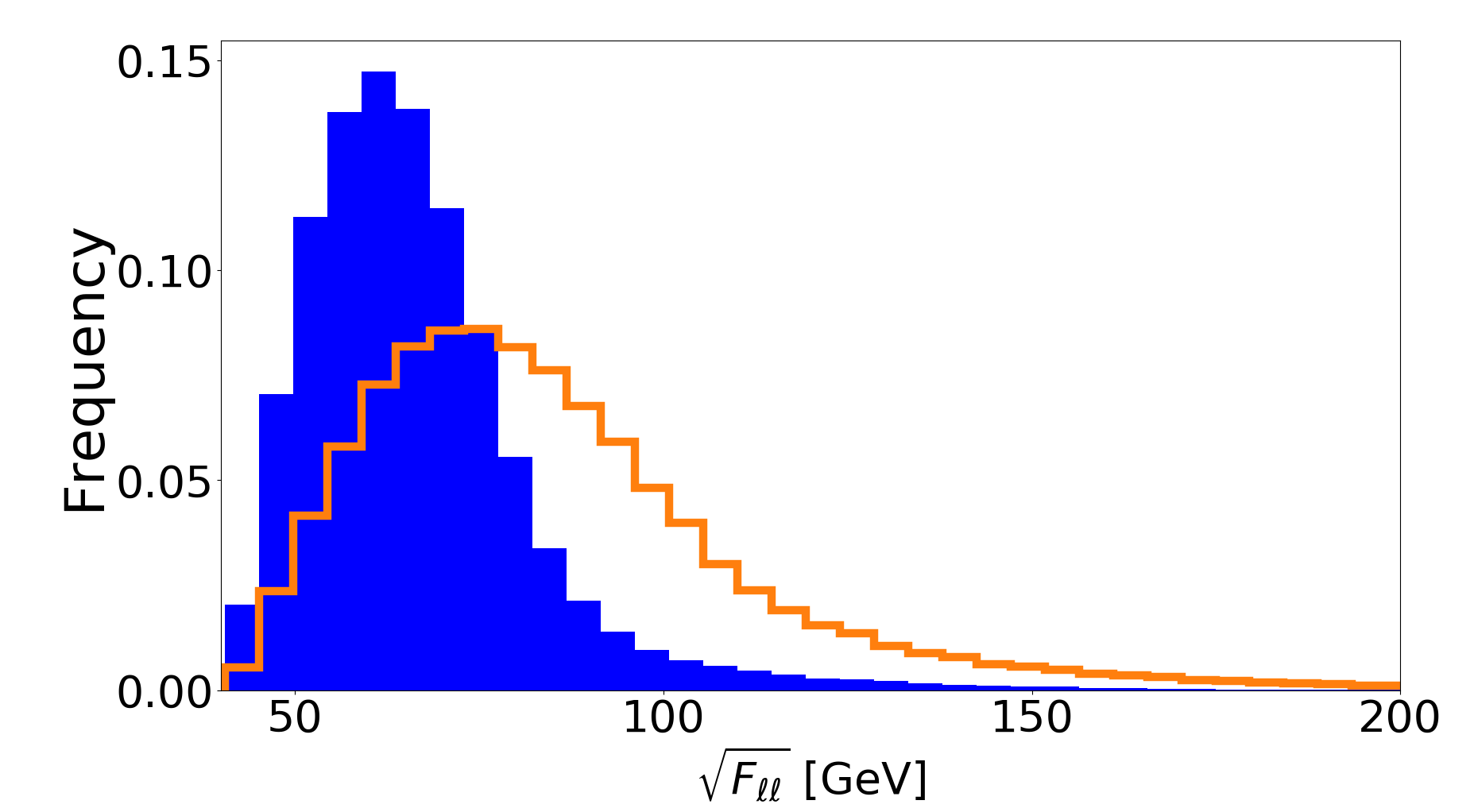}
		\label{Fsim}
	}
	\caption{Figures (a) and (b) show the dilepton mass $m_{\ell\ell}$ and $\sqrt{F_{\ell\ell}}$ distributions respectively for signal and background events.  }
	\label{constructed_mass}
\end{figure}

Figure \ref{AUC_with_constructedmass} shows the classification performance using a set of features comprising $\sqrt{F_{\ell\ell}}$ and six raw features, i.e. $\{\sqrt{F_{\ell\ell}},p_{T_1},\eta_1,\phi_1,p_{T_2},\eta_2,\phi_2 \}$ and another set comprising $m_{\ell\ell}$ and the same six raw features, i.e. $\{m_{\ell\ell},p_{T_1},\eta_1,\phi_1,p_{T_2},\eta_2,\phi_2 \}$ when using DNN and BDT. From the Figure, one can see that using $\sqrt{F_{\ell\ell}}$ as the only classification feature already performs the signal-background classification better than $m_{\ell\ell}$. Also, the classification performance curve in the same Figure did not change much when using $\{\sqrt{F_{\ell\ell}},p_{T_1},\eta_1,\phi_1,p_{T_2},\eta_2,\phi_2 \}$, which suggests that $\sqrt{F_{\ell\ell}}$ have utilized more relevant information from the raw features to separate the Higgs signal from the Z background compared to $m_{\ell\ell}$. The distributions of $\sqrt{F_{\ell\ell}}$ and $m_{\ell\ell}$ are shown in Figure \ref{constructed_mass}.

For the case of new physics searches where the masses of new particles are unknowns, one normally would not optimize the signal-background classifer based on a single set of signal parameters. Instead, one generates a range of possible mass values for the new physics, and construct new features which are hopefully not bias to any mass value. For such cases, constructing a good feature similar to $F_{\ell\ell}$ could still be done, but with the signal class being all events from the new physics with different possible parameters. In this manner, the DNN will be trained to construct signal-background separations that would put much more focus on the specific properties of the background process rather than on the specific properties of some new physics with specific parameters.

There are several observations that we can make on the coefficients $w_i$ of $F_{\ell\ell}$. The coefficients of the 3-momenta in $F_{\ell\ell}$, i.e. $w_x$,$w_y$ and $w_z$ are not all the same. In other words, $F_{\ell\ell}$ is not momenta-agnostic. In the case of $w_z$, this is understandable since in a detector environment, one already anticipates that the longitudinal momentum would not be similar to the transverse momenta. As for the $w_x$ and $w_y$ coefficients, they are the same to one significant figure. This suggests that there is no bias in the transverse momenta. We did a check on this with a convolutional neural network to find out if the transverse momenta $p_x$ and $p_y$ are indeed indistinguishable from one another. In this manner, the convolutional neural network (CNN) acts as a momentum bias checking tool.

\begin{figure}[h]
	\centering
	\subfigure[] {
		\includegraphics[width=0.46\textwidth]{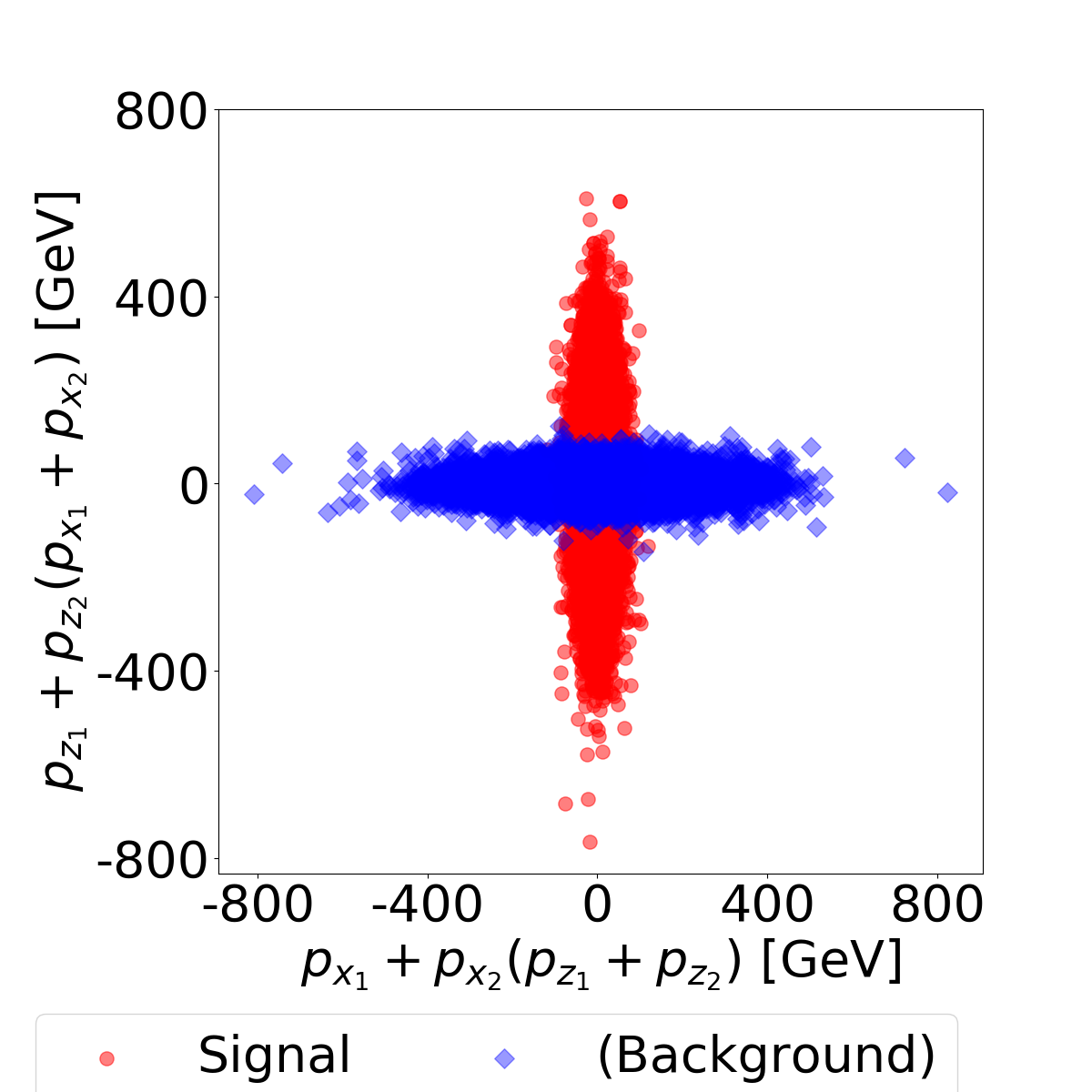}
		\label{pxpy_scatterplot}
	}
	\subfigure[] {
		\includegraphics[width=0.46\textwidth]{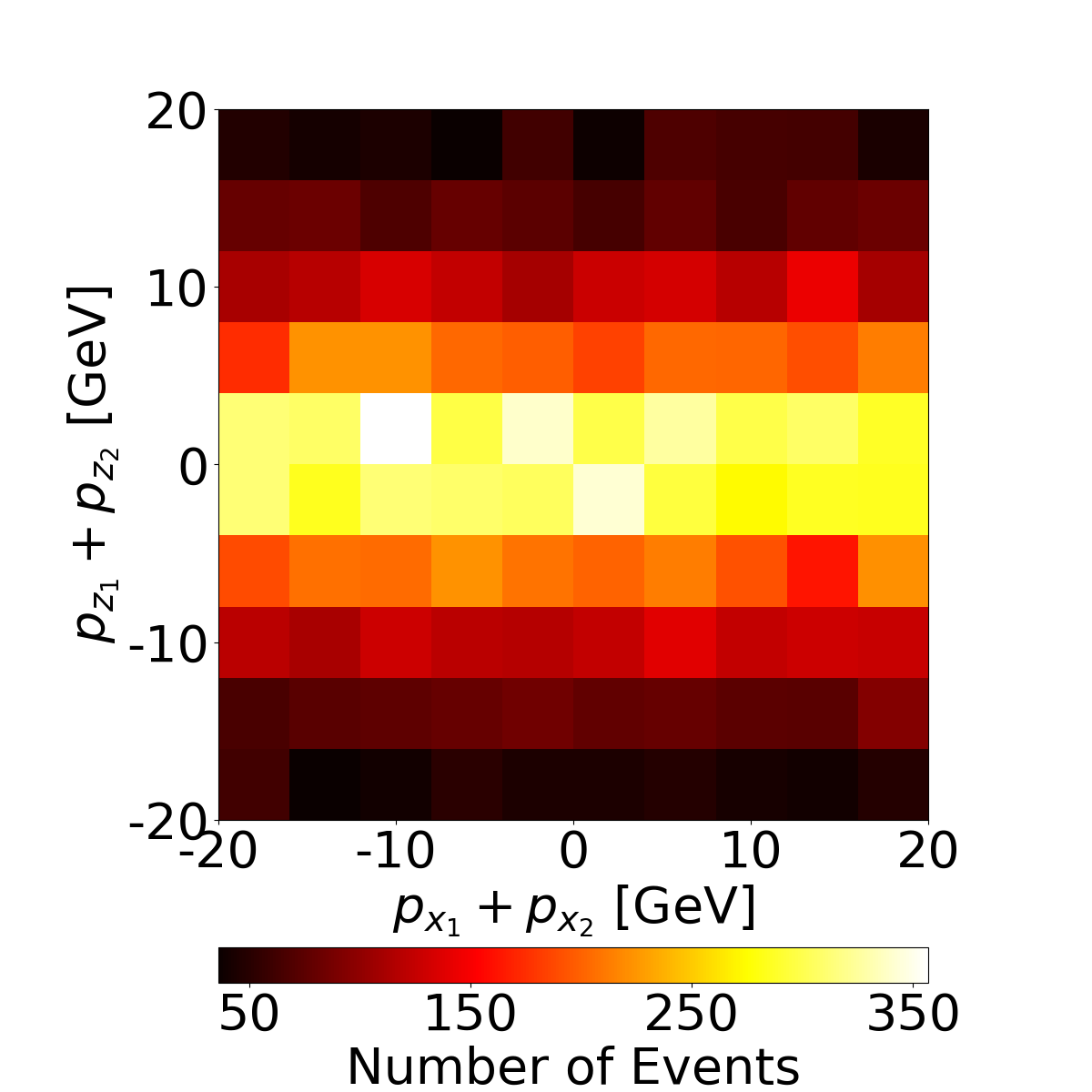}
		\label{pxpy_image}
	}
	\caption{Figure (a) shows a scatter plot of the signal $(p_{x_1} + p_{x_2})$ vs. $(p_{z_1} + p_{z_2})$ and background $(p_{z_1} + p_{z_2})$ vs. $(p_{x_1} + p_{x_2})$. Figure (b) shows an image of $(p_{x_1} + p_{x_2})$ vs. $(p_{z_1} + p_{z_2})$ converted from the scatter plot signal used in Figure (a).   }
	\label{pxpz}
\end{figure}

We first define our signal and background classes. Assuming the data points of $(p_{x_1} + p_{x_2})$ vs. $(p_{y_1} + p_{y_2})$ came from an unknown distribution, we define our signal to be scatter plots of  $(p_{x_1} + p_{x_2})$ vs. $(p_{y_1} + p_{y_2})$, and a swap of the axes as the background, i.e. the scatter plots of $(p_{y_1} + p_{y_2})$ vs. $(p_{x_1} + p_{x_2})$. A representative example of a signal and a background scatter plot is shown in Figure \ref{pxpy_scatterplot}.
\begin{figure}[h]
	\centering
	\subfigure[] {
		\includegraphics[width=0.46\textwidth]{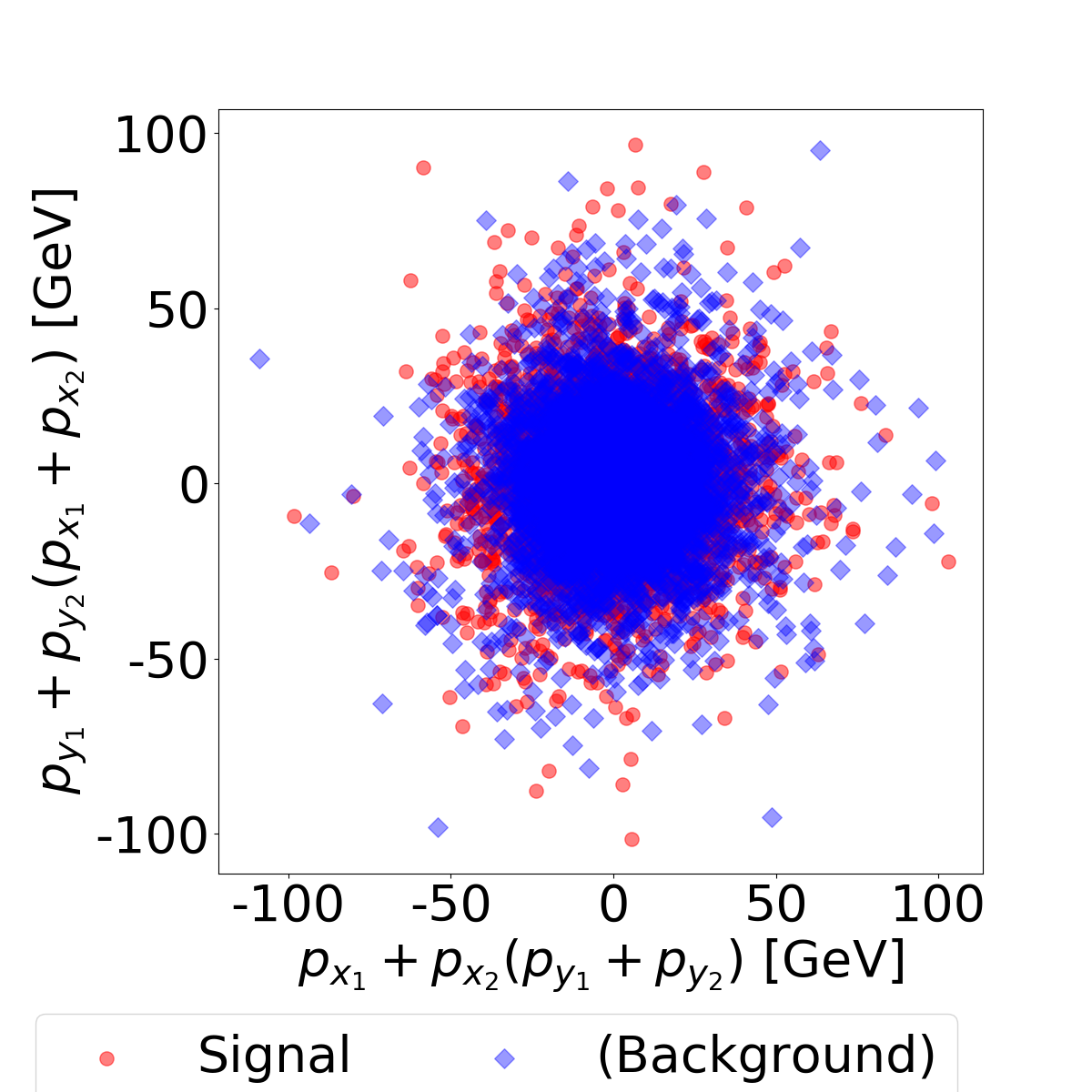}
		\label{pxpy_scatterplot}
	}
	\subfigure[] {
		\includegraphics[width=0.46\textwidth]{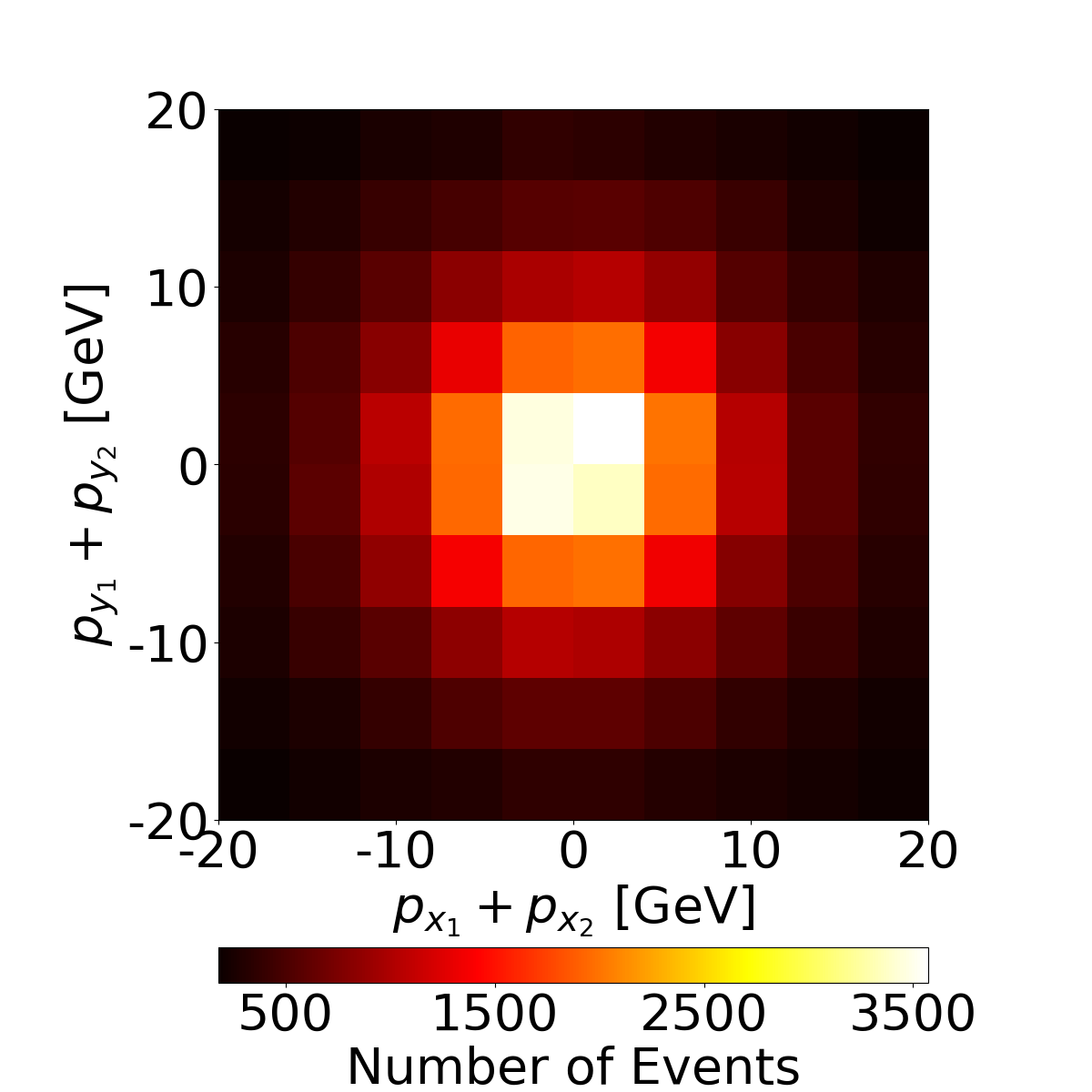}
		\label{pxpy_image}
	}
	\caption{Figure (a) shows a scatter plot of the signal $(p_{x_1} + p_{x_2})$ vs. $(p_{y_1} + p_{y_2})$ and background $(p_{y_1} + p_{y_2})$ vs. $(p_{x_1} + p_{x_2})$. Figure (b) shows an image of $(p_{x_1} + p_{x_2})$ vs. $(p_{y_1} + p_{y_2})$ converted from the scatter plot signal used in Figure (a).   }
	\label{pxpy}
\end{figure}
Each scatter plot contains the momenta from $N$ events randomly selected from the existing Monte Carlo dataset. In our work, we chose $N$ as 10 thousand. Since the inputs to the CNN have to be images, we converted each scatter plot to an image with 20 x 20 pixels (see Figure \ref{pxpy_image}).
Using a simple CNN with one convolution layer with 2x2 filters and two fully-connected layers, we find a classification accuracy of $0.5 \pm 0.006$ indicating that the CNN basically made random classifications, i.e. the CNN could find any bias in the transverse plane when comparing $p_x$ vs. $p_y$. Had if there was any bias, the CNN would be able to distinguish the signals from backgrounds. Rpeating this procedure with the signal being $(p_{x_1} + p_{x_2})$ vs. $(p_{z_1} + p_{z_2})$ and the background being $(p_{z_1} + p_{z_2})$ vs. $(p_{x_1} + p_{x_2})$ (see Figure \ref{pxpz}), we find a classification accuracy of $1$, indicating that the CNN regards the transverse momentum and longitudinal momentum as being entirely different from one another.


\section{Summary}

In this work, we studied on the importance of each feature relative to other classification features with deep learning and BDT. Two BSM processes have been used as case studies, namely multi-Higgs and SUSY scenario. From the feature importance ranking, one can determine the minimal number of features to obtain an optimal classification performance and discovery reach. Results indicate that the addition of features could still improve the discovery significance of the multi-Higgs case, but for the SUSY case, the significance has already plateau when using about ten features. We envisage the extension of this work to jet classifications; identifying the jet clusters that optimizes the classification or searching for clues on how best to increase the jet classification.

Moreover, in this work, the results show that the derived features in both multi-Higgs and SUSY scenario did not entirely capture the information content in the raw features. This motivates us to derive new classification features which could encapsulate more information content within the raw features, so that the raw features no longer have much impact on a physics discovery reach. In order to achieve this aim, we modified the DNN with a bottleneck neuron to construct new classification features which takes into account the detector effects. In this work, we demonstrated with a generic form of an invariant mass for the dilepton case from a Higgs $\to \tau\tau \to \ell\nu\ell\nu$ (as a signal) and $Z \to \tau\tau \to \ell\nu\ell\nu$ (as a background). We showed that our newly constructed feature performs better than a set of features that includes the dilepton invariant mass and the raw features of the two leptons, i.e. $\{m_{\ell\ell},p_{T_1},\eta_1,\phi_1,p_{T_2},\eta_2,\phi_2 \}$
containing the dilepton invariant mass $m_{\ell\ell}$ and the 3-momenta of the leptons in signal-background separations. Since the Higgs and Z decay to $\tau\tau$ have already been observed at the LHC, our feature construction approach can be readily tested in the experiments. It is straightforward to extend the feature construction approach to new physics searches.

As a side application of the feature construction approach, we could use the resulting constructed feature to identify any momentum biases. In this work, we have also used a convolutional neural network as part of the momentum bias checking approach. We performed a demonstration using the sum of the $p_x$ and $p_y$ momenta of the two leptons, and the sum of the $p_x$ and $p_z$ momenta of the same two leptons. In particular, the convolutional neural network gives an accuracy of 0.5 if two variables are statistically indistinguishable, and an accuracy of 1 if two variables are statistically distinguishable.






\acknowledgments

We wish to express our gratitude to Shen-Jian Chen and Zuo-Wei Liu for providing the computing facilities, including an Nvidia Tesla P40 GPU to complete this work. The authors gratefully acknowledge the support of the International Science and Technology Cooperation Program of China (Grant No. 2015DFG02100) and the National 973 Project Foundation of the Ministry of Science and Technology of China (Contract No. 2013CB834300).


\input{paper.bbl}

\end{document}

%% file: paper.bbl
\providecommand{\href}[2]{#2}\begingroup\raggedright\endgroup